\documentclass[aps,prd,showpacs,superscriptaddress,nofootinbib]{revtex4}
\usepackage{colordvi}

\usepackage{amsbsy}

\usepackage{amsfonts}

\usepackage{amssymb}

\usepackage{amsmath}

\usepackage{graphicx}
\usepackage{epsfig}

\def\ba{\begin{eqnarray}}
\def\ea{\end{eqnarray}}
\def\be{\begin{equation}}
\def\ee{\end{equation}}

\def\d{{\rm d}}

\def\={\mathrel{\widehat\mathalpha{=}}}
\def\puto#1{\rlap{\raise.5ex\hbox{\char'27}}{#1}}

\begin{document}





\title{Scalar hairy black holes and scalarons \\
in the isolated horizons formalism}

\date{November 8, 2005}




\author{Alejandro Corichi}
\email{corichi@nucleares.unam.mx} \affiliation{Instituto de
Ciencias
Nucleares, Universidad Nacional Aut\'onoma de M\'exico,\\
A. Postal 70-543, M\'exico D.F. 04510, M\'exico}
\affiliation{Instituto de Matem\'aticas, Universidad Nacional
Aut\'onoma de M\'exico\\ A. Postal 61-3, Morelia, Michoac\'{a}n, 58090, M\'exico}

\author{Ulises Nucamendi}
\email{ulises@itzel.ifm.umich.mx} \affiliation{Instituto de
F\'{\i}sica y Matem\'{a}ticas, Universidad Michoacana de San
Nicol\'{a}s de Hidalgo,
Edif. C-3, Ciudad Universitaria, Morelia, Michoac\'{a}n,  58040,
M\'{e}xico}

\author{Marcelo Salgado}
\email{marcelo@nucleares.unam.mx} \affiliation{Instituto de
Ciencias Nucleares, Universidad Nacional Aut\'onoma de M\'exico,\\
A. Postal 70-543, M\'exico D.F. 04510, M\'exico}




\begin{abstract}
The Isolated Horizons (IH) formalism, together with a simple
phenomenological model for colored black holes has been used to
predict non-trivial formulae that relate the ADM mass of the
solitons and hairy Black Holes of Gravity-Matter system on the one
hand, and several horizon properties of the black holes in the
other. In this article, the IH formalism is tested numerically for
spherically symmetric solutions to an Einstein-Higgs system where
hairy black holes were recently found to exist.  It is shown that
the mass formulae still hold and that, by appropriately extending
the current model, one can account for the behavior of the horizon
properties of these new solutions. An empirical formula that
approximates the ADM mass of hairy solutions is put forward, and
some of its properties are analyzed.

\end{abstract}

 \maketitle





\section{Introduction}

\bigskip

\noindent In recent years, the introduction of the Isolated
Horizon (IH) formalism \cite{PRL,Ashtekar:2000yj,AFK} has proved
to be useful to gain insight into the static sector of theories
admitting ``hair'' \cite{CS,ACS}.  Firstly, it has been found that
the Horizon Mass of the black hole (BH), a notion constructed out
of purely quasi-local quantities, is related in a simple way to
the ADM masses of both the colored black hole and the solitons of
the theory \cite{CS}. Second, a simple model for colored black
holes as bound states of regular black holes and solitons has
allowed to provide heuristic explanations for the behavior of
horizon quantities of those black holes \cite{ACS}. Third, the
formalism is appropriate for the formulation of {\it uniqueness
conjectures} for the existence of unique stationary solutions in
terms of horizon ``charges" \cite{CS}. Finally, the combination of
the Mass formula, together with the fact that in theories such as
Einstein-Yang-Mills-Higgs (EYMH) different ``branches'' of static
solutions merge, has allowed to have a formula for the difference
of soliton masses in terms of black hole quantities
\cite{Kleihaus:2000kv,ACS,CNS}. Many of these predictions have
been confirmed in more general situations and for other matter
couplings \cite{horizon,horizon2}. For a recent review on IH
(including hair) see \cite{AK2}, and for a review of hairy black
holes see \cite{volkov}.

In this article, we explore further the consequences of the IH
formalism in the static sector of the theory. In particular, we
explore the behavior of a recently found family of hairy static
spherically symmetric (SSS) solutions to the Einstein-Higgs system
\cite{SN}, where the scalar potential is allowed to be negative
and therefore, the existing no-hair theorems \cite{hair} do not
apply. In the standard treatment of stationary black holes with
killing horizons, one is always restoring to several concepts that
use asymptotic information very strongly  \cite{HeuslerB}. On the
other hand, the IH formalism only uses quasi-local information
defined on the horizon, allowing it to prove very general results
involving only these quasi-local quantities. The IH formalism has
proved to be generalizable, in the scalar sector, even to the
non-minimal coupling regime, where the energy conditions required
for the consistency of the formalism are much weaker \cite{ACS2}.
In the present paper we shall restrict our attention to the
minimally coupled case, and for a particular form of the scalar
potential for which static solutions are known to exist \cite{SN}.
We will study the one parameter family of solutions (that could be
labelled by its geometric radius $r_\Delta$) and compare its
properties with those of hairy black holes in other theories, such
as EYM, where the phenomenological predictions of the IH formalism
have been shown to work very well \cite{ACS,horizon}. As we will
show, we find that the mass formulae relating BH and soliton ADM
masses works also well, but the model of a hairy black hole as a
bound state of a soliton and a bare black holes exhibits some new
unexpected features. As we  shall see, one needs to slightly
modify the model from its original formulation in Ref.\cite{ACS}. Once
this modification is made the model can again explain all the
qualitative behavior of the hairy BH solutions.

The structure of the paper is as follows: In Sec.~\ref{sec:2} we
review the consequences of the IH formalism for hairy solitons and
BH solutions. In Sec.~\ref{sec:3} we review the SSS found recently
in the Einstein-Higgs sector. Section~\ref{sec:4} is the main
section of the paper. In it, we show the numerical evidence for
the mass formulae and the phenomenological predictions of the
model. Unlike the EYM case, there are some unexpected features,
such as the binding energy becoming positive. We then propose a
modification of the formalism to deal with such situations. We
show that with these modifications, the model can still account
for the geometrical phenomena found in several theories. In
Sec.~\ref{sec:5} we explore the situation of the collapse of a
hairy black hole and use the model to put bound on the total
possible energy to be radiated. These results should be of some
relevance to full dynamical numerical evolutions of such black
holes. In Sec.~\ref{sec:6} we propose an empirical formula for the
horizon and ADM masses of scalar hairy black holes that can also
be applied to the EYM case. Finally, we end with a discussion in
Sec.~\ref{sec:7}.


\section{Consequences of the Isolated horizons formalism}
\label{sec:2}
\bigskip





\noindent In recent years, a new framework tailored to consider
situations in which the black hole is in equilibrium (``nothing
falls in''), but which allows for the exterior region to be
dynamical, has been developed. This {\it Isolated Horizons} (IH)
formalism is now in the position of serving as starting point for
several applications. Notably, for the extraction of physical
quantities in numerical relativity and also for quantum entropy
calculations \cite{PRL, AK2}. The basic idea is to consider
space-times with an interior boundary (to represent the horizon),
satisfying quasi-local boundary conditions ensuring that the
horizon remains `isolated'. Although the boundary conditions are
motivated by geometric considerations, they lead to a well defined
action principle and Hamiltonian framework. Furthermore, the
boundary conditions imply that certain `quasi-local charges',
defined at the horizon, remain constant `in time', and can thus be
regarded as the analogous of the global charges defined at
infinity in the asymptotically flat context. The isolated horizons
Hamiltonian framework allows to define the notion of {\it Horizon
Mass} $M_\Delta$, as a function of the `horizon charges'
(hereafter, the subscript ``$\Delta$" stands for a quantity at the
horizon).

In the Einstein-Maxwell and Einstein-Maxwell-Dilaton systems
considered originally \cite{Ashtekar:2000yj}, the horizon mass
satisfies a Smarr-type formula and a generalized first law in
terms of quantities defined exclusively at the horizon (i.e.
without any reference to infinity). The introduction of non-linear
matter fields like the Yang-Mills field has brought unexpected
subtleties to the formalism~\cite{CS}. However, one still is in
the position of defining a Horizon Mass, and furthermore, this
Horizon Mass satisfies a first law.

An isolated horizon is a non-expanding null surface generated by a
(null) vector field $l^a$. The IH boundary conditions imply that
the acceleration $\kappa$ of $l^a$ ($l^a\nabla_al^b=\kappa l^b$)
is constant on the horizon $\Delta$. However, the precise value it
takes on each point of phase space (PS) is not determined
a-priori. On the other hand, it is known that for each vector
field $t^a_o$ on space-time, the induced vector field $X_{t_{o}}$
on phase space is Hamiltonian if and only if there exists a
function $E_{t_{o}}$ such that $\delta E_{t_{o}}=\Omega
(\delta,X_{t_{o}})$, {\it for any vector field $\delta$ on PS}.
This condition can be re-written
 as, $
\delta E_{t_{o}}=\frac{\kappa_{t_{o}}}{8\pi G}\,\delta a_{\Delta}
+ {\rm work\;\; terms}$. Thus, the first law arises as a necessary
and sufficient condition for the consistency of the Hamiltonian
formulation. Thus, the allowed vector fields $t^a$ will be those
for which the first law holds. Note that there are as many `first
laws' as allowed vector fields $l^a\widehat{=} \,t^a$ on the
horizon. However, one would like to have a {\it Physical First
Law}, where the Hamiltonian $E_{t_{o}}$ be identified with the
`physical mass' $M_{\Delta}$ of the horizon. This amounts to
finding the `right $\kappa$'. This `normalization problem' can be
easily overcome in the EM system \cite{Ashtekar:2000yj}. In this
case, one chooses the function $\kappa=\kappa(a_\Delta, Q_\Delta)$
as the corresponding function for the {\it static} solution with
charges $(a_\Delta, Q_\Delta)$. However, for the EYM system, this
procedure is not as straightforward. A consistent viewpoint is to
abandon  the notion of a globally defined  horizon mass on Phase
Space, and to define, for each value of $n=n_o$ (which labels
different branches of the solutions), a canonical normalization
$t^a_{n_o}$ that yields the Horizon Mass $M^{(n_o)}_{\Delta}$ for
the $n_o$ branch \cite{CS,AFK}. The horizon mass takes the form
(from now on we shall omit the $n_0$ label),
\begin{equation}
M_{\Delta}(r_\Delta)=\frac{1}{2G_0}\int_0^{r_\Delta} \beta(r) \,
\d r\, ,
\end{equation}
with $r_\Delta$ the horizon radius. Here $\beta(r_{\Delta})$ is
related to the surface gravity as follows
$\beta(r_{\Delta})=2r_{\Delta}\kappa(r_{\Delta})$.

Furthermore, one can relate the horizon mass $M_{\Delta}$ to the
ADM mass of static black holes.  Recall first that general
Hamiltonian considerations imply that the total Hamiltonian,
consisting of a term at infinity, the ADM mass, and a term at the
horizon, the Horizon Mass, is constant on every connected
component of static solutions (provided the evolution vector field
$t_0^a$ agrees with the static Killing field everywhere on this
connected component) \cite{Ashtekar:2000yj,AFK}. In the
Einstein-Yang-Mills case, since the Hamiltonian is constant on any
branch, we can evaluate it at the solution with zero horizon area.
This is just the soliton, for which the horizon area $a_\Delta$,
and the horizon mass $M_{\Delta}$ vanish.  Hence we have that
$H^{(t_0)} = M_{\rm sol}$. Thus, we conclude \cite{CS}:
\begin{equation}
\label{ymmass} M_{\rm sol} = M_{\rm ADM} - M_{\Delta}\, ,
\end{equation}
Thus, the ADM mass contains two contributions, one attributed to
the black hole horizon and the other to the outside `hair',
captured by the `solitonic residue'. The formula (\ref{ymmass}),
together with some energetic considerations \cite{ACS}, lead to
the model of a colored black hole as a bound state of an ordinary,
`bare', black hole and a `solitonic residue', where the ADM mass
of the colored black hole of radius $r_{\Delta}$ is given by the
ADM mass of the soliton plus the horizon mass of the `bare' black
hole plus the binding energy:
\begin{equation}
\label{ymbinding}
   M_{\rm ADM} = M_{\rm sol} +
M_{\Delta}= M^0_\Delta +  M_{\rm sol} + E_{\rm bind}\,\,\,,
\end{equation}
with $E_{\rm bind}= M_{\Delta} - M^0_\Delta$. Simple
considerations about the behavior of the ADM masses of the colored
black holes and the solitons, together with some expectations of
this model (such as demanding for a non-positive binding energy)
give raise to several predictions about the behavior of the
horizon parameters \cite{ACS}. Among the predictions, we have:

i) The absolute value of the binding energy decreases as
$r_\Delta$ increases.

ii) $\beta(r_{\Delta})$, as a function of $r_\Delta$, is a
positive function, bounded above by $\beta_{(0)}(r_{\Delta})=1$.

iii) The curve $\beta(r)$, as functions of $r$ intersect the $r=0$
axis at distinct points between $0$ and $1$, and never intersect.
Finally,

iv) The curve for $\beta$, for large value of its argument,
becomes asymptotically tangential to the curve
$\beta_{(0)}(r_\Delta)=1$.

One of the features of these solutions in, say, EYM is that there
is no limit for the size of the black hole. That is, if we plot
the ADM mass of the BH as function of the radius $r_\Delta$ we get
an infinite number of curves, each of the intersecting the
$r_{\Delta}=0$ line at the value of the soliton mass, and never
intersecting each other.
\bigskip

The purpose of this paper is to test the mass formula
(\ref{ymmass}), for the scalar hairy solutions found in
Ref.~\cite{SN} and also to confront the predictions i)-iv)
(obtained for the colored EYM BH model \cite{ACS}), with the
corresponding properties for the scalar hairy BH. In the next
section we will review the scalar hairy solutions, and afterwards
we shall study their horizon properties.





\section{Scalar solitons and black holes in Einstein-Higgs theory}
\label{sec:3}
\bigskip

\noindent Let us consider the theory of a scalar field minimally
coupled to gravity described by the total action:
\begin{equation}
S_{\rm tot}[g_{\mu\nu},\phi]
 = \int \sqrt{-g}\left\{ \frac{R}{16\pi} -
\left[\frac{1}{2}
(\nabla^{\beta}\phi)(\nabla_{\beta} \phi) + V(\phi) \right]\right\}
{\rm d}^4 x
\label{acttot}
\end{equation}
(units where $G_0=c=1$ are employed).
The field equations following from the
variation of the action (\ref{acttot}) are,
\begin{equation}
\label{eqsmov} G_{\mu\nu} = 8\pi
\left\{(\nabla_\mu\phi)\nabla_\nu\phi - g_{\mu\nu}\left[\frac{1}{2}
(\nabla^{\beta}\phi) (\nabla_{\beta} \phi) + V(\phi)\right]
\right\} \ ,
\end{equation}
and,
\begin{eqnarray}
\Box \phi = {\frac{\partial V(\phi)}{\partial \phi}} \ .
\label{eqssca}
\end{eqnarray}

\noindent It is well known that asymptotically flat static
spherically symmetric solutions representing black holes solutions
to the Einstein-Higgs equations do not exist if the scalar matter
satisfies the weak energy condition (WEC) due to the existence of
the so called scalar no-hair theorems \cite{hair}. Recently,
numerical evidence for the existence of asymptotically flat and
static spherically symmetric solutions representing scalar hairy
black holes (SHBH) and scalar solitons ({\it scalarons}; hereafter
SS) have been found in theories represented by the action
(\ref{acttot}) and with a scalar potential
non-positive-semidefinite \cite{SN} given by the asymmetric
potential,
\begin{eqnarray}
\label{potential} V(\phi) &=& \frac{\lambda}{4}\left[ (\phi-a)^2 -
\frac{4(\eta_1 + \eta_2)}{3} (\phi-a) +  2\eta_1\eta_2 \right]
(\phi-a)^2 \,\,\,,
\end{eqnarray}
where $\lambda$, $\eta_i$ and $a$ are constants. For this class of
potential one can see that, for $\eta_1>2\eta_2>0$, $\phi=a$
corresponds to the local minimum, $\phi=a+\eta_1$ is a global
minimum and $\phi=a+\eta_2$ is a local maximum (see Fig. 1). The
key point in the shape of the potential, $V(\phi)$, for the
asymptotically flat solutions to exist, is that the local minimum
$V^{\rm local}_{\rm min}=V(a)$ is also a zero of $V(\phi)$ (to see
\cite{SN} for an detailed analysis for the existence of these
solutions). Moreover, $V(\phi)$ is not positive definite (we
assume $\lambda>0$), which leads to a violation of the WEC and
therefore the scalar no-hair theorems \cite{hair} can not be
applied to this case.

\begin{figure}[h]
\centerline{ \epsfig{figure=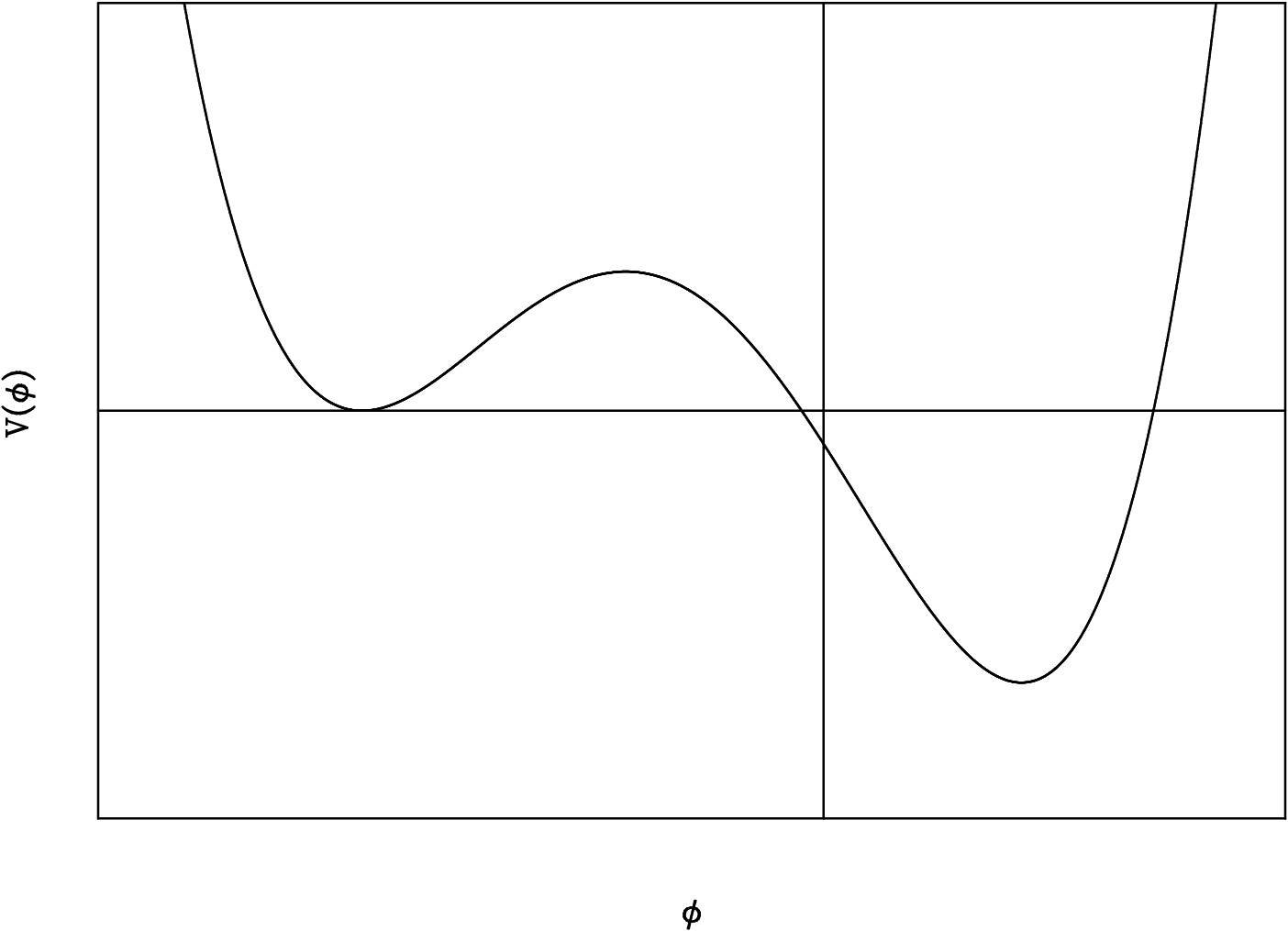,width=5.2cm,angle=-90}} \vspace*{0.5cm}
\caption{} Qualitative shape of the scalar-field potential $V(\phi)$
as given by Eq. (\ref{potential}) used to construct the asymptotically flat
black hole and soliton solutions. \label{fig0}
\end{figure}

In order to describe the asymptotically flat SHBH and SS, we use a
standard parametrization for the metric and the scalar field
describing spherically symmetric and static spacetimes
\begin{eqnarray}
ds^2 &=& - \left(1-\frac{2m(r)}{r}\right) e^{2\delta(r)} dt^2 +
\left(1-\frac{2m(r)}{r}\right)^{-1} dr^2 + r^2 d\Omega^2
\,\,\,\,\,,
\label{metric} \\
\phi &=& \phi(r) \,\,\,\,\,, \label{scalar}
\end{eqnarray}

For SHBH we demand regularity on the event horizon $r_{\Delta}$
which implies the conditions,
\begin{equation}
m_{\Delta} = \frac{r_{\Delta}}{2}, \,\,\,\,\,\, \delta(r_{\Delta})
= \delta_{\Delta}, \,\,\,\,\,\, \phi(r_{\Delta}) = \phi_{\Delta},
\label{EHconditions1}
\end{equation}

\begin{equation}
(\partial_{r}\phi)_{\Delta} =
\frac{r_{\Delta}(\partial_{\phi}V)_{\Delta}}{(1 - 8\pi
r_{\Delta}^2 V_{\Delta})} \,\,\,, \,\,\,\,\,\,
(\partial_{r}m)_{\Delta} = 4\pi r_{\Delta}^2 V_{\Delta} \,.
\label{EHconditions2}
\end{equation}
For SS we impose regularity at the origin of coordinates $r=0$,
\begin{equation}
m(0) = 0, \,\,\,\,\,\, \delta(0) = \delta_{0},
\,\,\,\,\,\, \phi(0) = \phi_{0},\,\,\,\,\,\,
(\partial_{r}\phi)_{0} = 0 \,. \label{origencond}
\end{equation}
where $\delta_{0}$ and $\phi_{0}$ are to be found such as to
obtain the desired asymptotic conditions. In addition to the
regularity conditions, we impose asymptotically flat conditions on
the spacetime for SHBH and SS:
\begin{equation}
m(\infty) = M_{\rm ADM}, \,\,\,\,\,\, \delta(\infty) = 0,
\,\,\,\,\,\, \phi(\infty) = \phi_{\infty} \,\,\,\,\,\, .
\label{asympcond}
\end{equation}
Above, the value $\phi_{\infty}$ corresponds to the local minimum
of $V(\phi)$. $M_{\rm ADM}$ is the ADM mass associated with a SHBH
or SS configuration. For a given theory, the family of SHBH
configurations is parametrized by the free parameter $A_{\Delta}$
which specifies the area of the black hole horizon. Therefore for
SHBH, $M_{\rm ADM}=M_{\rm ADM}(A_{\Delta})$, or equivalently
$M_{\rm ADM}=M_{\rm ADM}(r_{\Delta})$ since in our coordinates the
horizon area $A_{\Delta}=4\pi r_{\Delta}^2$. The value
$\phi_{\Delta}=\phi_{\Delta}(r_{\Delta})$ is a shooting parameter
rather than an arbitrary boundary value which is determined so
that the asymptotic flat conditions are satisfied. On the other
hand, for SS the value $\phi_0$ is the shooting parameter, and the
corresponding configuration is characterized by a unique $M_{\rm
ADM}^{\rm sol}$.

Finally, the surface gravity of a spherically symmetric static black
hole can be calculated from the general expression of spacetimes
admitting a Killing horizon \cite{HeuslerB}:
\begin{equation}
\kappa = \left[ -\frac{1}{4}\nabla^2 \alpha \right]^{1/2}_{r=r_{\Delta}}
\,\,\,\,\,\,, \label{defsg0}
\end{equation}
where $\nabla^2$ stands for the Laplacian operator associated with the stationary
metric and $\alpha= (\partial_t,\partial_t)$ is the norm of the time-like
(static) Killing field which is null at the horizon.
For the present case, $\alpha= g_{tt}=
- \left(1-\frac{2m(r)}{r}\right) e^{2\delta(r)}$.

From the above formula, one can obtain the following useful expression
\begin{equation}
\kappa = \lim_{r \rightarrow r_{\Delta}} \left\{\frac{1}{2}
\frac{\partial_r g_{tt}}{\sqrt{g_{tt}g_{rr}}} \right\}
\,\,\,\,\,\, . \label{defsg}
\end{equation}
For the election of the parametrization of the metric
(\ref{metric}) we have
\begin{equation}
\kappa (r_{\Delta}) = \frac{1}{2r_{\Delta}}
e^{\delta(r_{\Delta})} \left[ 1 - 2(\partial_{r}m)_{\Delta}
\right] \,\,\,\,\,\, . \label{sg}
\end{equation}
Introducing (\ref{EHconditions2}) in (\ref{sg}) we obtain the
final expression for the surface gravity of the SHBH
\begin{equation}
\kappa (r_{\Delta}) = \frac{1}{2r_{\Delta}}
e^{\delta(r_{\Delta})} \left[ 1 - 8\pi r_{\Delta}^2 V_{\Delta}
\right] \,\,\,\,\,\, . \label{sgSHBH}
\end{equation}
In the next section, we shall analyze these solutions from the
perspective of the IH formalism.





\section{Mass formulae}
\label{sec:4}
\bigskip

\begin{figure}[h]
\centerline{
\epsfig{figure=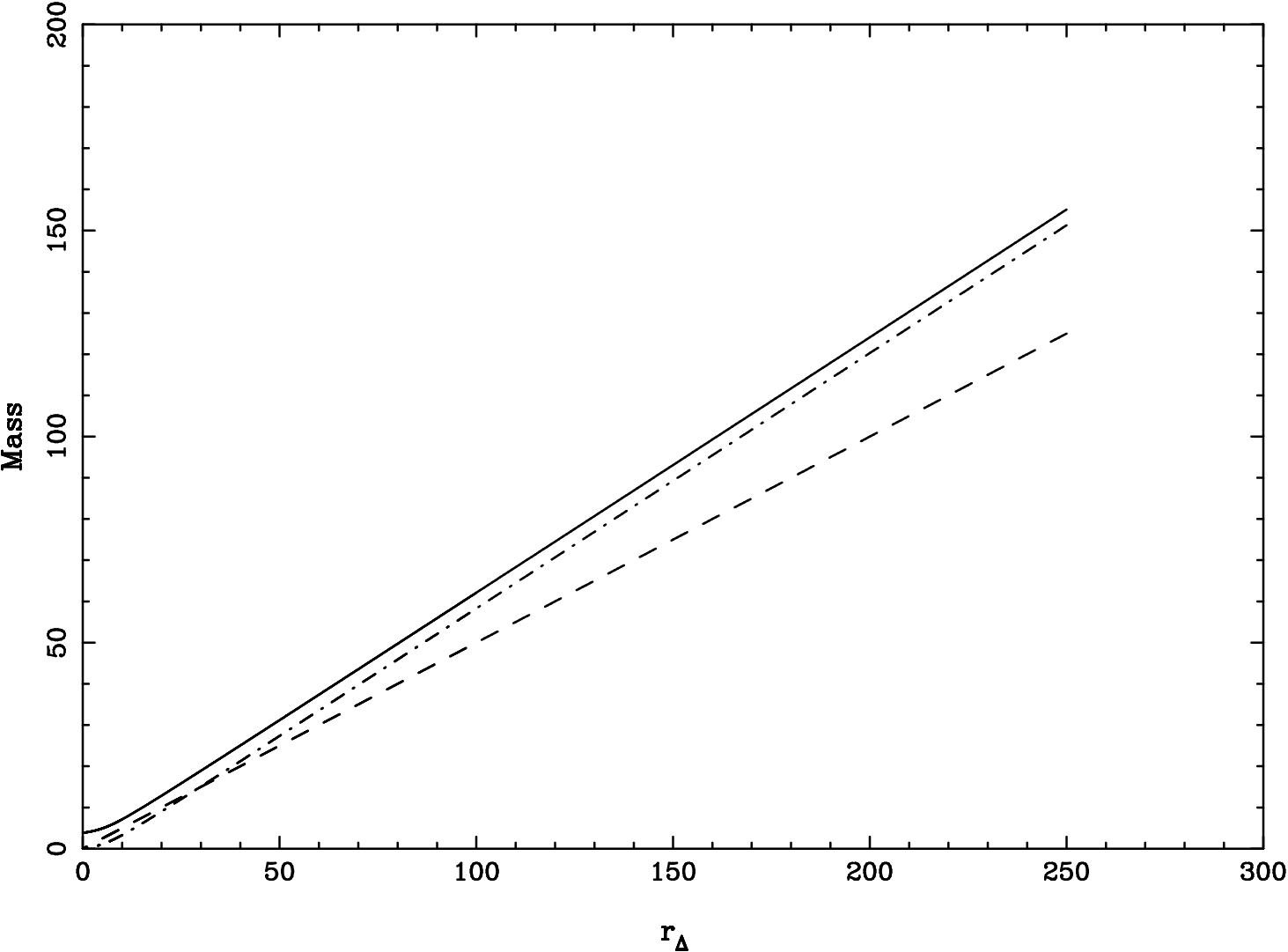,width=6.5cm,angle=-90}
\epsfig{figure=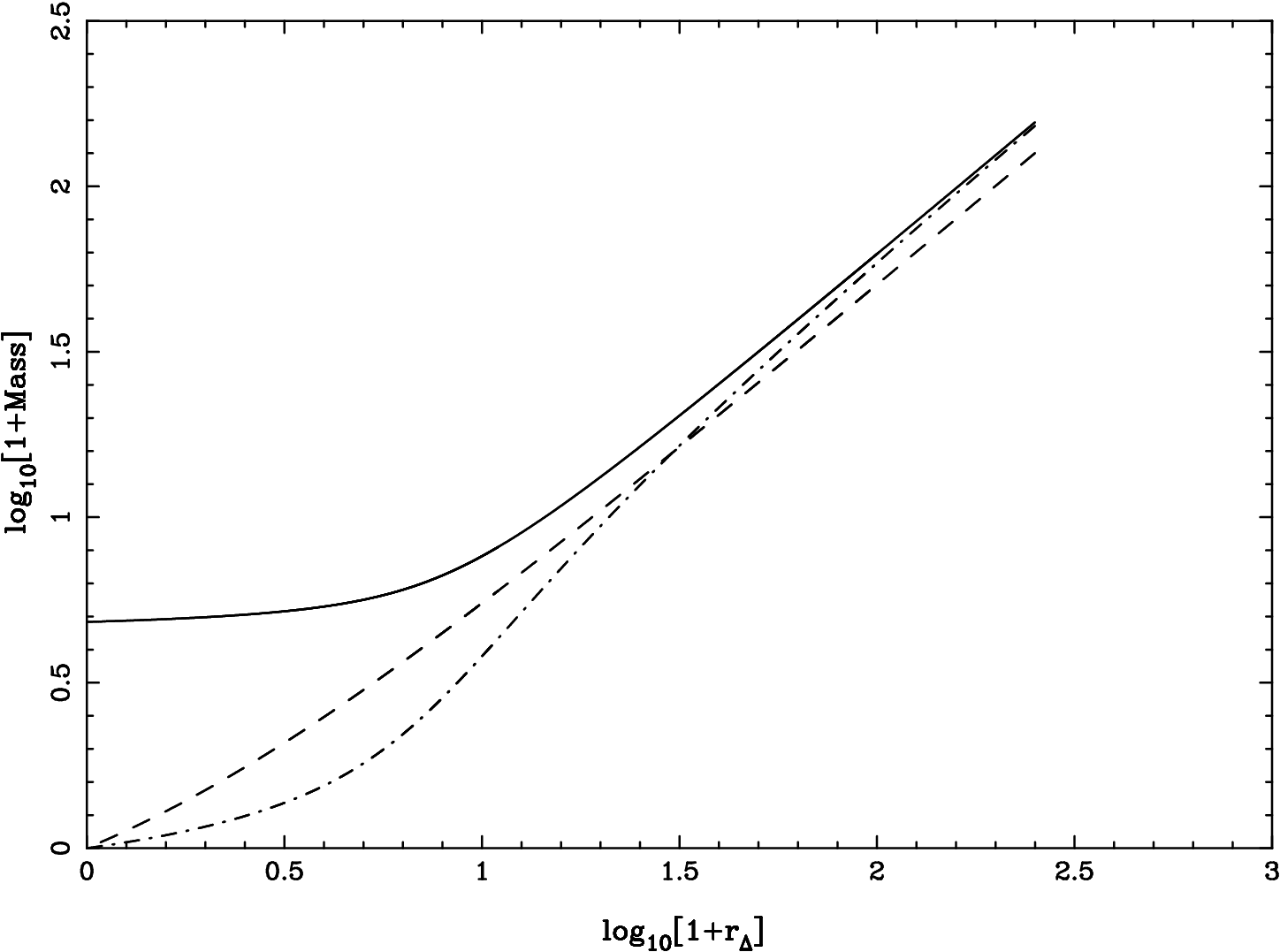,width=6.5cm,angle=-90}
}
\vspace*{0.5cm}
\caption{}
The ADM mass (solid lines), the horizon BH mass (dash-dotted lines),
and the mass of the Schwarzschild solution (dashed-lines)
plotted as functions of $r_\Delta$ (first panel).
The second panel depicts similar quantities using logarithmic scales
to appreciate better their behavior for small $r_\Delta$.
The soliton mass $M_{\rm sol}\approx 3.827$.
\label{masses}
\end{figure}

\begin{figure}[h]
\centerline{
\epsfig{figure=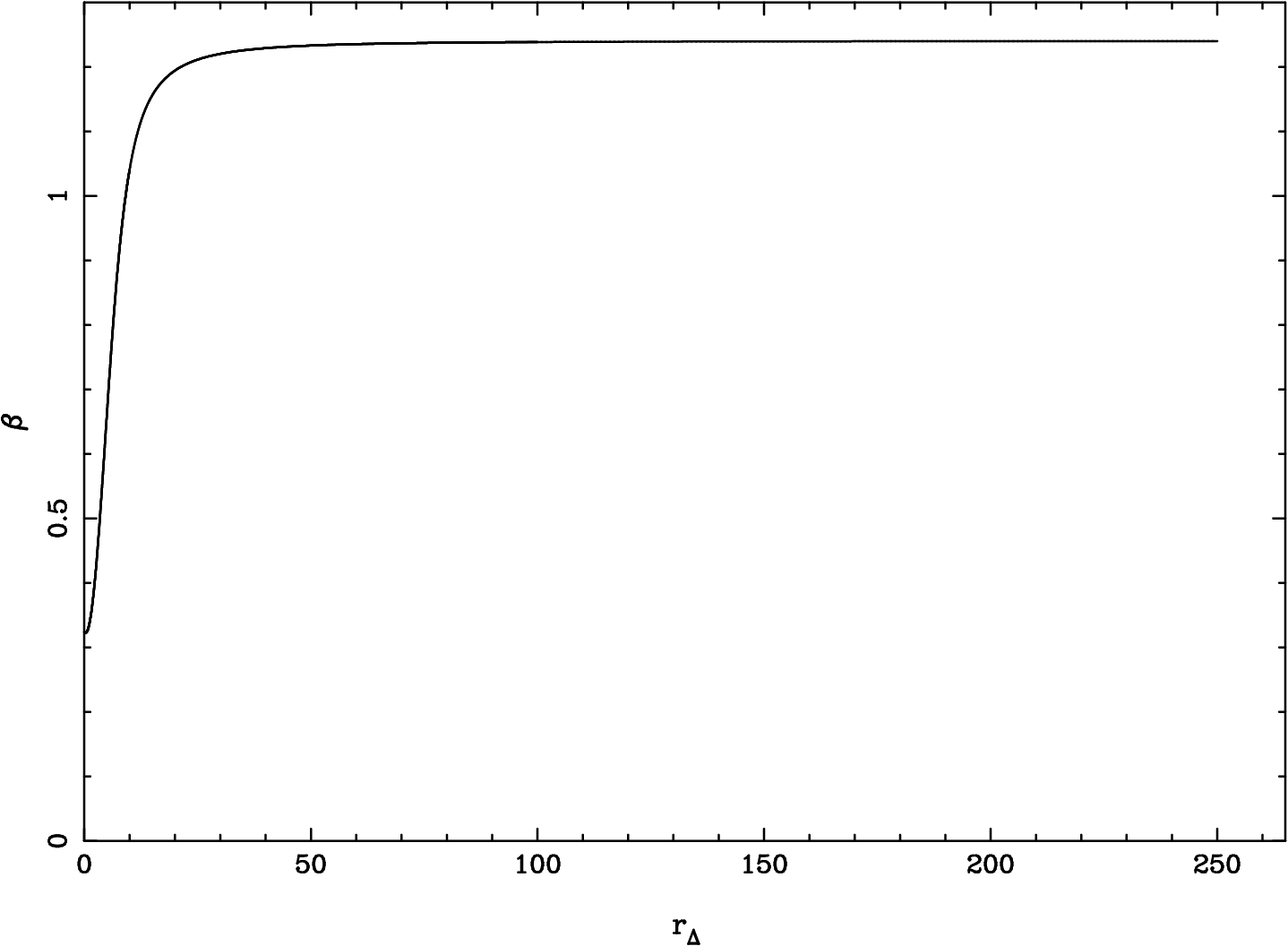,width=7cm,angle=-90}
}
\vspace*{0.5cm}
\caption{}
$\beta(r_\Delta)=2\kappa\,r_\Delta$ is
  plotted as function of  $r_\Delta$. Note that it approaches
  asymptotically the value $\approx 1.24$. Here $\beta(0)\approx 0.324$.
\label{beta}
\end{figure}

\noindent
 Let us now turn to the straightforward application of
the IH formalism mentioned in the Sec. II to the case of SHBH and
SS in the Einstein-Higgs theory with action given by
(\ref{acttot}) and $V(\phi)$ by Eq. (\ref{potential}) \cite{SN}.
As in Ref. \cite{SN}, we shall take the specific values
$\eta_1=0.5$, $\eta_2=0.1$ and $a=0$; all the quantities (e.g.,
$M_{\rm ADM}$ and $r_\Delta$) have been rescaled as appropriate
using $1/\sqrt{\lambda}$ as a length-unit.

 The first consequence coming
from the IH formalism is that the horizon mass associated to the
SHBH takes the form\footnote{We remind the reader our choice of units
$G_0=c=1$.},
\begin{equation}
\label{MH}
M_{\Delta}(r_\Delta)=\frac{1}{2}\int_0^{r_\Delta}\beta(r) \, \d
r\, ,
\end{equation}
where $\beta(r_{\Delta})=2r_{\Delta}\kappa(r_{\Delta})$ [the value
of $\kappa (r_{\Delta})$ is given by (\ref{sgSHBH})]. We have
dropped the subindex $(n)$ in the expression (\ref{MH}) because in
the Einstein-Higgs system considered here there is only one branch
of static spherically symmetric SHBH labelled by its horizon
radius $r_\Delta$ (the corresponding scalar configurations do not
have nodes). Additionally, there is another branch of static
spherically symmetric BH given by the family of Schwarzschild BH's
labelled by its corresponding horizon radius $r_\Delta$ and with
horizon mass
\begin{equation}
\label{MHS} M_{\Delta}^{\rm Schwarz}(r_\Delta)=
r_\Delta/2 \,\,\,\,.
\end{equation}
The second consequence coming from the IH
formalism is that on the entire branch of SHBH we can expect the
following identity to be true
\begin{equation}
\label{EHmass}
M_{\rm ADM}= M_{\rm sol} + M_{\Delta} \,\,\,,
\end{equation}
where $M_{\rm sol}$ is the ADM mass of the SS obtained taking the
limit $r_\Delta \rightarrow 0$ of the branch of SHBH and $M_{\rm
ADM}$ is the ADM mass corresponding to the SHBH with horizon radius
$r_\Delta$, and the horizon mass is given by (\ref{MH}). Thus, in a
similar way to the EYM theory, the total ADM Mass of the solution
contains two contributions, one attributed to the horizon of the
SHBH and the other to the outside `hair', captured by the SS. We
have performed numerical explorations for SHBH for a large range of
values of the horizon radius (in normalized units) and have checked
the identity (\ref{EHmass}). We have found complete agreement within
the numerical uncertainties. This can be seen in Fig.~\ref{masses},
where the identity was checked up to $r_{\Delta}=250$.

Figure~\ref{beta} depicts the behavior of $\beta(r_{\Delta})$.
Unlike the EYM model, where $\beta\approx 1$ for large $r_{\Delta}$,
in this model $\beta\approx 1.24$ asymptotically.

\begin{figure}[h]
\centerline{ \epsfig{figure=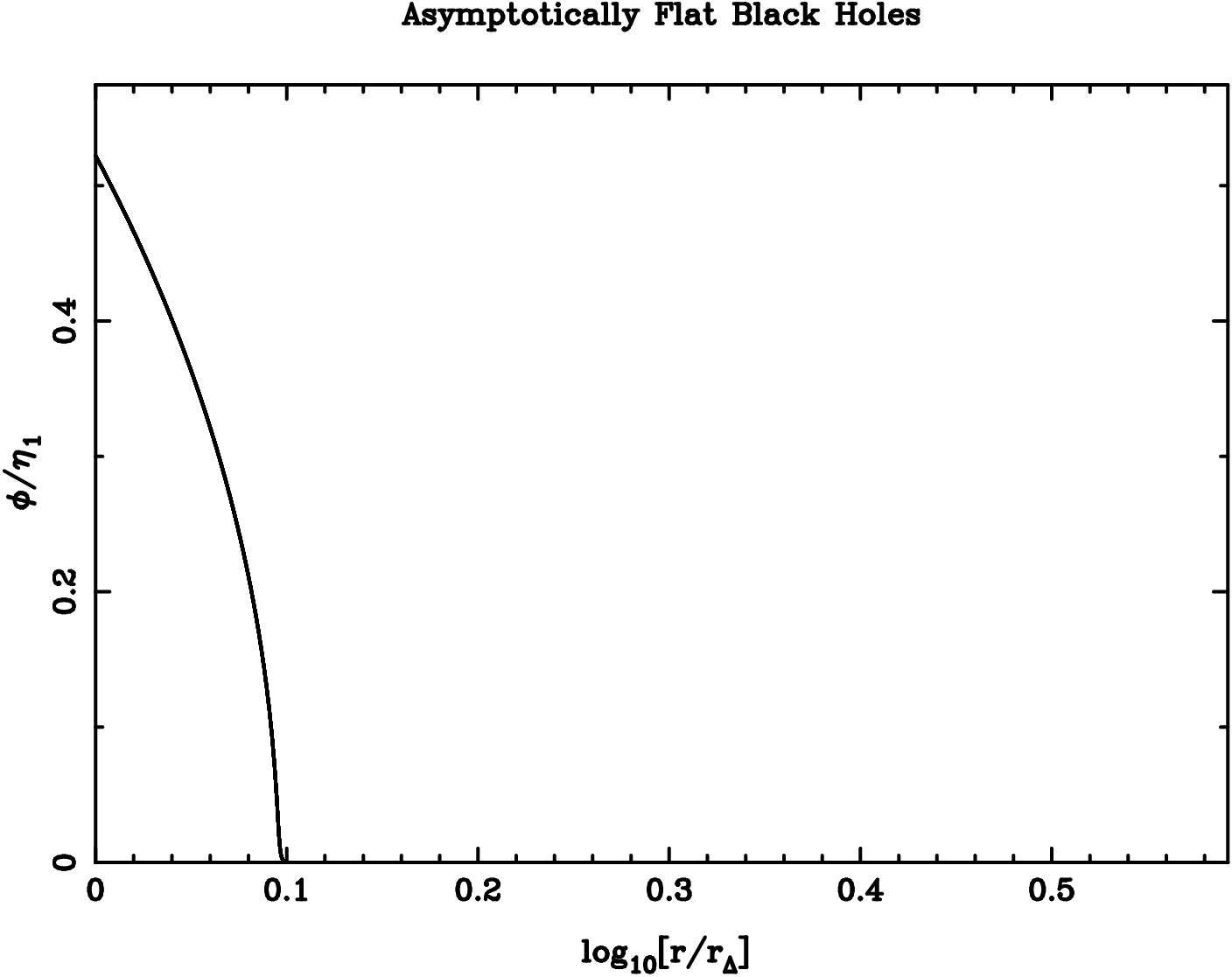,width=5.2cm,angle=-90}
\epsfig{figure=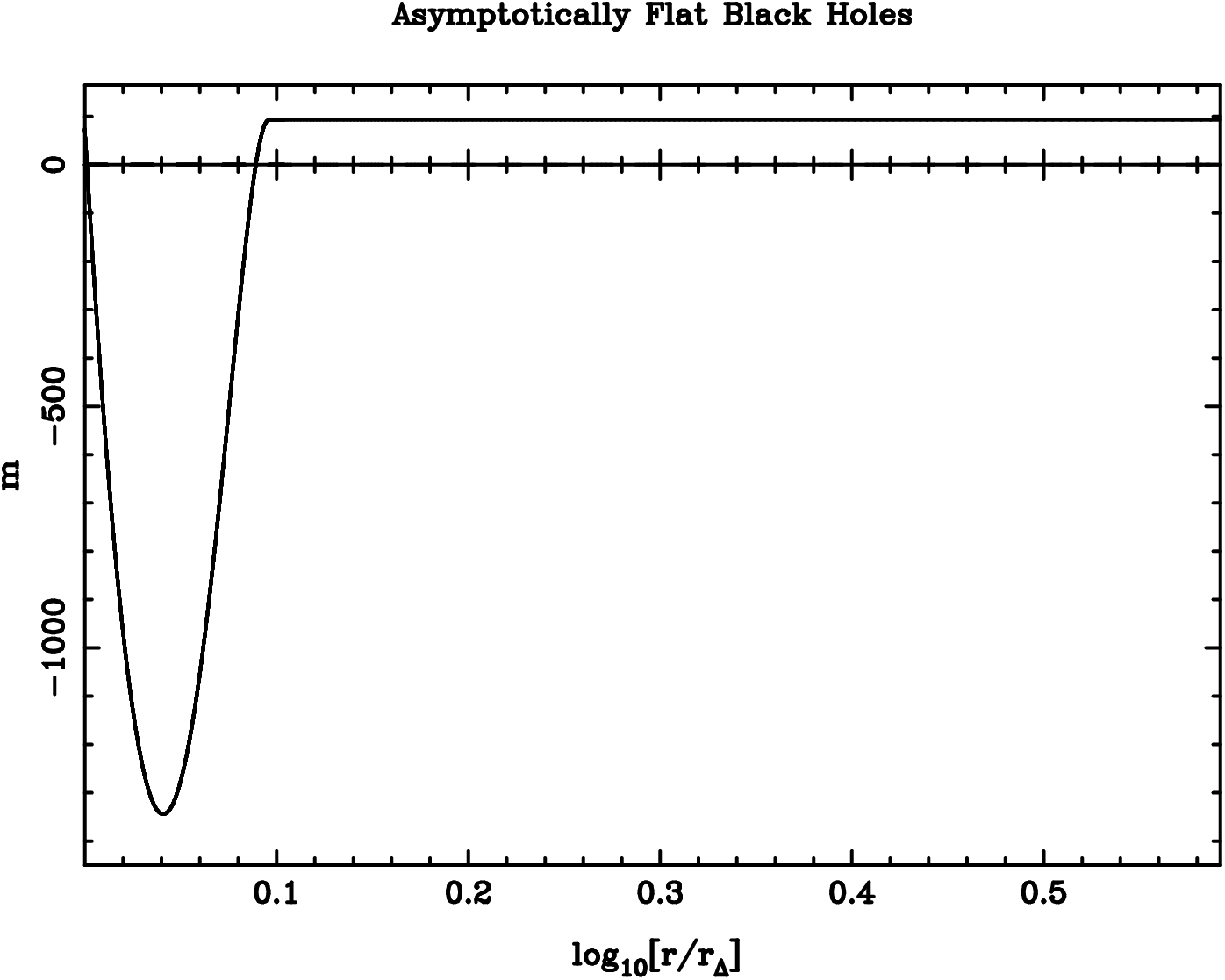,width=5.2cm,angle=-90} }
\vspace*{0.3cm} \centerline{
\epsfig{figure=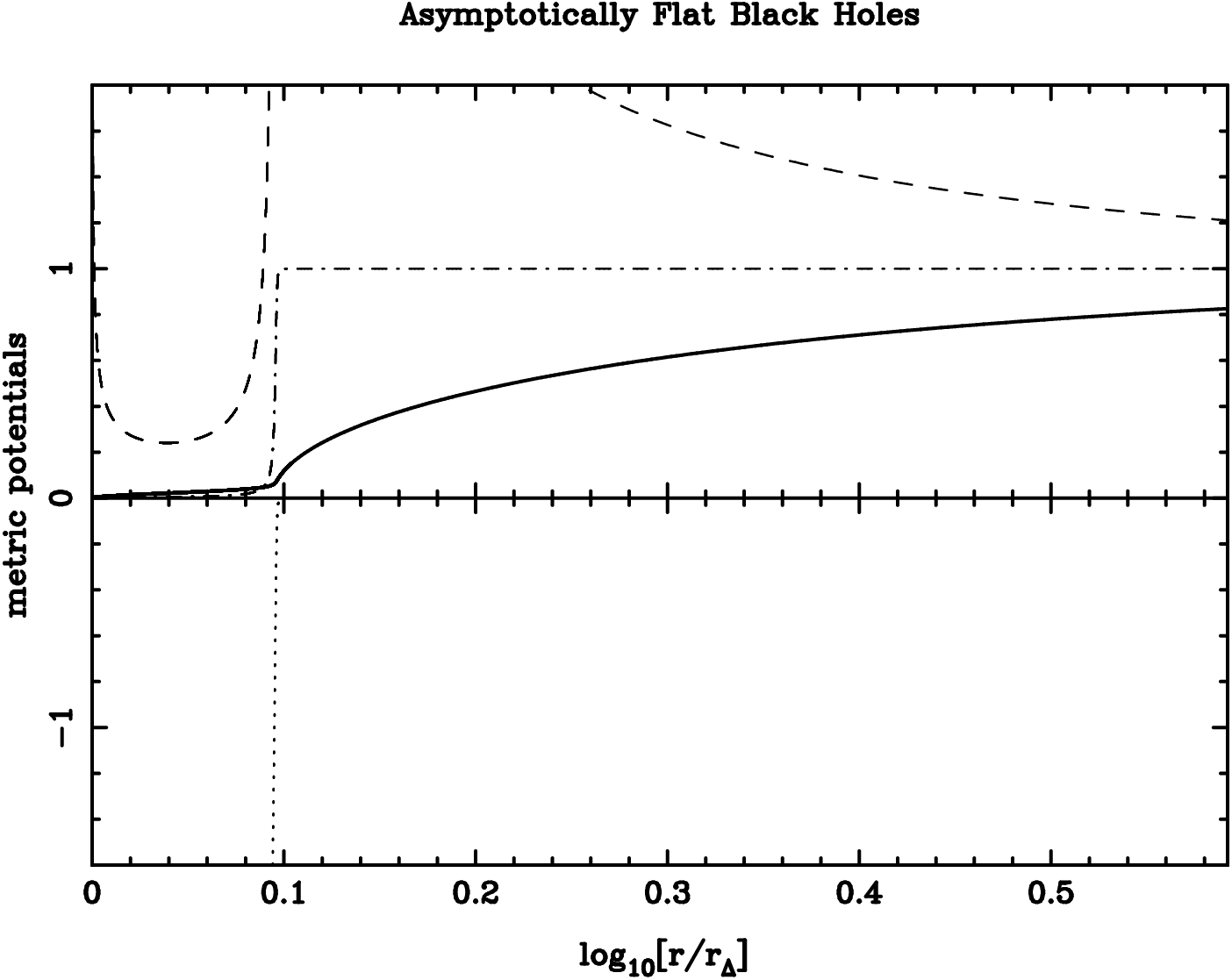,width=5.2cm,angle=-90}
\epsfig{figure=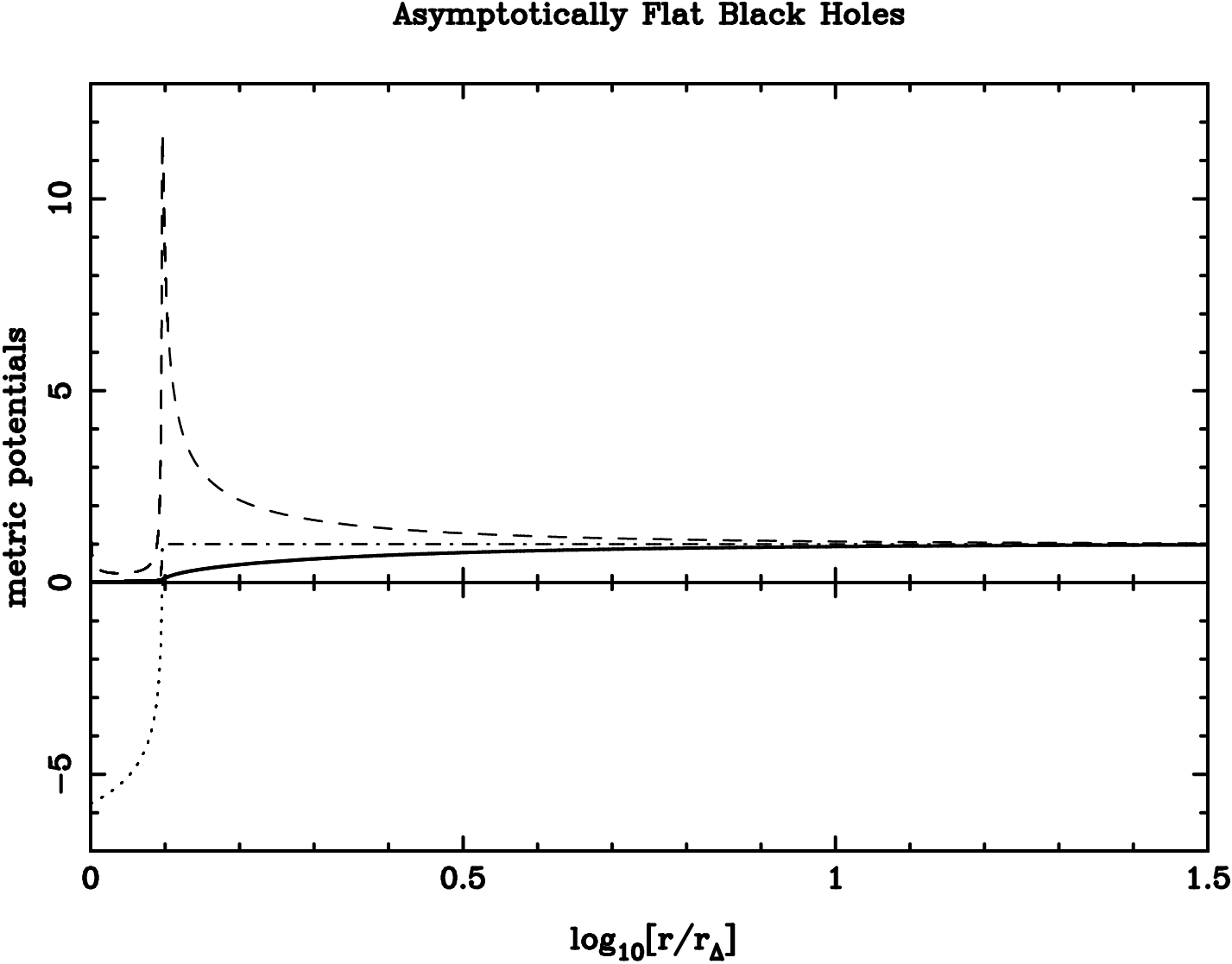,width=5.2cm,angle=-90} }
\vspace*{0.5cm} \caption{} Large-black-hole configuration
constructed with $V(\phi)$ as given by Eq. (\ref{potential}) with
parameters $\eta_1= 0.5$, $\eta_2= 0.1$, $a= 0$, and
$r_{\Delta}=150/\sqrt{\lambda}$, $\phi_{\Delta}\sim 0.26111$. The
upper panels depict the scalar field and the mass function
respectively. The latter converges to $M_{\rm ADM}\sim 93.096
/\sqrt{\lambda}$. The lower panels depicts the metric potentials
(the first is a zoom of the second): $\sqrt{-g_{tt}}$ (solid
line), $\sqrt{g_{rr}}$ (dashed line), $e^{\delta}$ (dash-dotted
line) and $\delta$ (dotted line). \label{BHconf}
\end{figure}

Figure~\ref{BHconf} shows an example of a BH solution with large
$r_{\Delta}$ (for a small BH see fig.2 of Ref. \cite{SN}).

\subsection{A physical model of SHBH}
\bigskip

\noindent The extrapolation of the model of a hairy black hole as
a bound state of an ordinary, `bare', black hole and a `solitonic
residue' (first applied successfully to the colored BH's in the
EYM theory) \cite{ACS} does not apply directly to the SHBH because
the straightforward generalization of the formula
(\ref{ymbinding}) as
\begin{equation}
\label{PM}
M_{\rm ADM} = M_{\rm sol} + M_{\Delta}= M^{\rm
schwarz}_\Delta + M_{\rm sol} + E_{\rm bind}\,\,\,\,,
\end{equation}
where $E_{\rm bind}= M_{\Delta} - M^{\rm schwarz}_\Delta$, to our
case, has the problem that the binding energy changes sign,
becoming positive for BH larger that $r_{\Delta}\approx 30$, and then
increasing in absolute value as $r_\Delta$ gets larger. That is,
\begin{equation}
E_{\rm bind} \sim B r_\Delta  \,\,\,\,,
\end{equation}
(where $B$ is a constant whose value depends on the specific model) 
for $r_\Delta \gg 1$, which is contrary to the
expected feature of a negative binding energy as in the EYM case
(i.e. the prediction (i) mentioned above will not be satisfied).

In order to appreciate the origin of the failure of this feature,
let us recall that we can write $M_{\rm ADM}$ in terms of the
Schwarzschild mass and the mass of the `hair' as,
\begin{equation}
\label{FM}
M_{\rm ADM}(r_\Delta) = M^{\rm
schwarz}_\Delta(r_\Delta) + M_{\rm hair}(r_\Delta) \,\,\,,
\end{equation}
where
\begin{equation}
M_{\rm hair}(r_\Delta) = - \int^{\infty}_{r_\Delta} r^2 {T}^{t}_{t}
\d r \,\,\,\,, \label{mhairint0}
\end{equation}
By rescaling the $r-$coordinate in terms of $r_\Delta$, we can
rewrite the mass of the hair. It becomes then,
\begin{equation}
M_{\rm hair}(r_\Delta) = - r_\Delta \int^{\infty}_{1} x^2
\tilde{T}^{t}_{t} \d x \,\,\,\,, \label{mhairint}
\end{equation}
with
\begin{eqnarray}
 x &=& \frac{r}{r_\Delta}\,\,\,,\\
 \tilde{T}^{t}_{t} &=& r_\Delta^2 T^{t}_{t} \,\,\,,\\
 T^{t}_{t} &=& - \left[ \left(1-\frac{2m(r)}{r}\right)
\frac{(\partial_r \phi)^2}{2} + V(\phi) \right]\,\,\,.
\end{eqnarray}

Now, for $x\gg 1$, the integral in Eq.(\ref{mhairint}) becomes almost
independent of $r_\Delta$, and in fact the numerical analysis provides
the following value
\begin{equation}
 M_{\rm hair}(r_\Delta\gg 1) \sim B\, r_{\Delta} \,\,\,.
\end{equation}
where $B\approx 0.12$.
Now, since $M^{\rm
schwarz}_\Delta(r_\Delta)= r_\Delta/2$ we have then
\begin{equation}
M_{\rm ADM}(r_\Delta\gg 1) \sim C r_\Delta \,\,\,.
\end{equation}
where now $C\approx 0.62$. Therefore we conjecture that $C$ is a
constant that depends of the matter-theory involved. For the EYM
case, one can easily show that the scaling properties of the hair
contribution of the energy makes the equivalent of the integral of
Eq.(\ref{mhairint}) to behave like $1/r_\Delta$ rather than
$r_\Delta$. Therefore, for the EYM case, $C=1/2$. As we now show,
this subtle difference in both theories makes that the binding
energy expression used in the EYM cannot be used straightforwardly
 for the Einstein-Higgs theory analyzed here.

 From (\ref{PM}) and (\ref{FM}) we find that
\begin{equation}
E_{\rm bind}(r_\Delta\gg 1) \sim M_{\rm
hair}(r_\Delta\gg 1) - M_{\rm sol} \sim B\, r_\Delta - M_{\rm sol},
\end{equation}
then $E_{\rm bind}$ scales as $r_\Delta$ when $r_\Delta \gg 1$ (remember that $M_{\rm sol}$ is a constant) \cite{comment}.
It is clear that the sign of the binding energy as such defined
changes sign and grows with the size of the black hole.

To deal with this situation we now proceed to propose a
modification of the model of a hairy black hole as a bound state
in order to adapt it to the more general case. Our proposal
consists in  redefining the binding energy as,
\begin{equation}
E_{\rm bind}^{\rm new}(r_\Delta) = E_{\rm bind}(r_\Delta) -
B\,r_\Delta = M_{\Delta}(r_\Delta) - M^{\rm
schwarz}_\Delta(r_\Delta) - B\,r_{\Delta}\, ,
\end{equation}
By this procedure we have `renormalized' the binding energy by
subtracting the divergent term.
That is, the new expression for the binding energy is
\begin{equation}
\label{ebindnew}
E_{\rm bind}^{\rm new}(r_\Delta) =M_{\Delta}(r_\Delta) - C\,r_{\Delta}\, .
\end{equation}
This definition shares now exactly the same properties as the
original expression for the EYM. Thus, it vanishes at
$r_\Delta=0$, and decreases monotonically to the negative value
$-M_{\rm sol}$. It is interesting to note that this new definition
reduces to the old one for the EYM case, since as we previously
remarked $C_{\rm EYM}=1/2$.

\begin{figure}[h]
\centerline{ \epsfig{figure=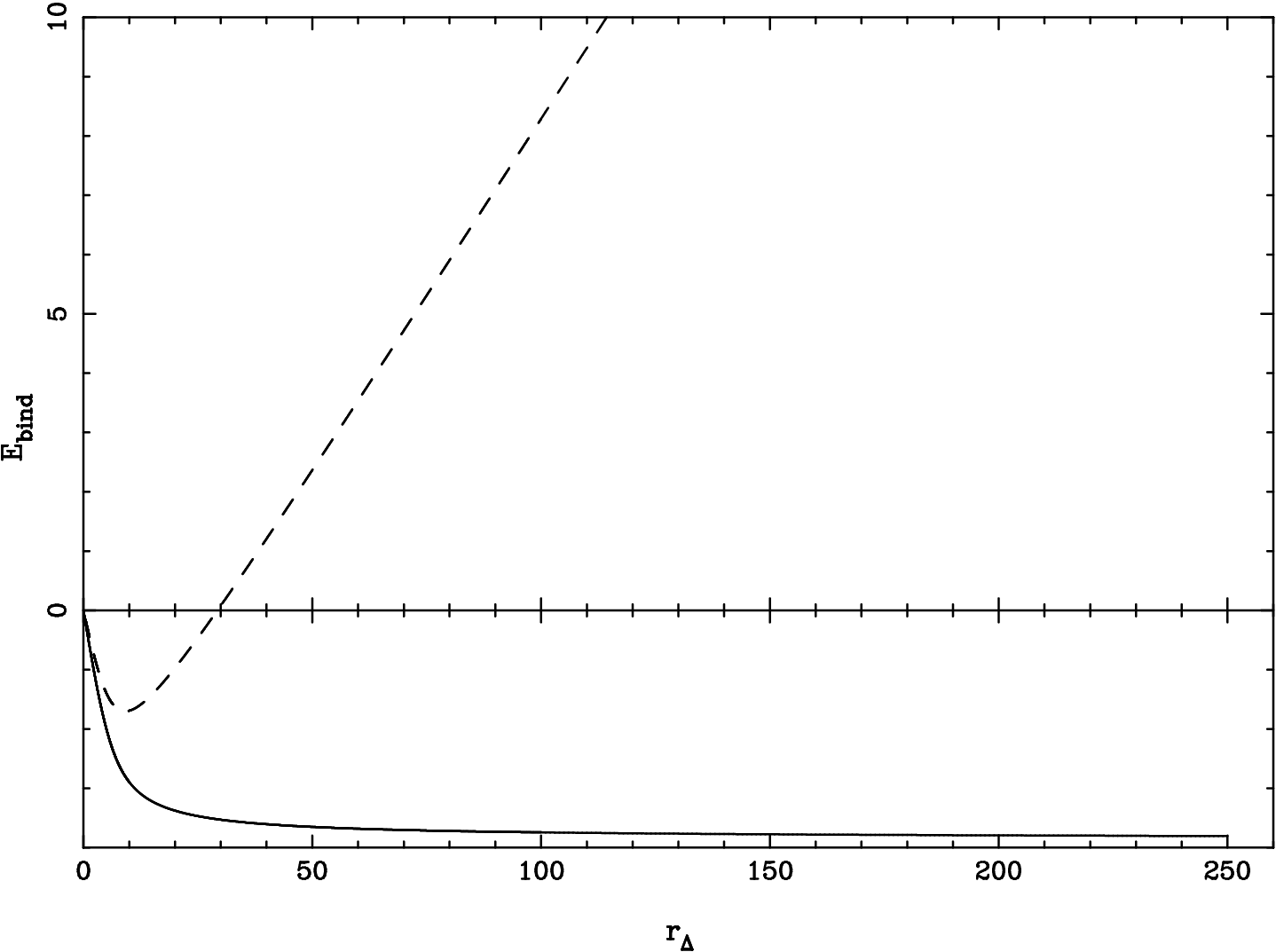,width=7cm,angle=-90} }
\vspace*{0.5cm} \caption{} Both binding energies are plotted. The
`old' binding energy $E_{\rm bind}(r_\Delta)$ is shown to become
positive and approaches asymptotically a straight line. The `new'
binding energy $E_{\rm bind}^{\rm new}(r_\Delta)$  is also plotted
(continuous line), showing the expected behavior. \label{bindE}
\end{figure}

The formula (\ref{PM}) can now be reformulated as,
\ba \label{PMnew} M_{\rm ADM}(r_\Delta) &=& M_{\rm sol} +
M_{\Delta}(r_\Delta)\\ \nonumber &=& M^{\rm schwarz}_\Delta
(r_\Delta)+ M_{\rm sol} + E_{\rm bind}^{\rm new}(r_\Delta) + B\,
r_\Delta\\ \nonumber & = & C\,r_{\Delta} + M_{\rm sol} + E_{\rm
bin}^{\rm new}(r_\Delta),
\ea Thus, we would get the same structure for the ADM mass of the
hairy black hole where now the mass of the `bare black hole' would
be equal to $\tilde{M}^0_\Delta= (1/2+B) r_\Delta= C r_\Delta$.
However it is not clear what the origin of this extra term
($Br_\Delta$) is, and we could very well have assigned it to the
soliton mass to form a new `solitonic residue´ with mass $M_{\rm
sol}+B r_\Delta$ (whose interpretation however seems somewhat
obscure). We must explore more in order to decide which
interpretation is best suited for our model. As we stressed, the
constant $C$ depends on the theory considered; for the EYM and
EYMH theories, $C=1/2$. It would be interesting to explore (in
addition to the current analysis, see below) whether other
theories admitting hair posses values of $C$ different from $1/2$.

To end this section, let us rewrite the form of the hairy mass in
terms of the old binding energy, in order to understand the
behavior of the scalar system. First, let us note using Eqs.
(\ref{PM}) and (\ref{FM}) that
\begin{equation}
\label{gg}
M_{\rm hair}(r_\Delta)= M_{\rm ADM}(r_\Delta)- M^{\rm
schwarz}_\Delta(r_\Delta)= M_{\rm sol}+E_{\rm bind} \,\,\,,
\end{equation}
so the binding energy is
\begin{equation}
\label{be} E_{\rm bind}=M_{\rm hair}(r_\Delta)-M_{\rm sol}= -\left[
\int^{\infty}_{r_\Delta} r^2 {T}^{t}_{t} \d r -\int^{\infty}_{0} r^2
{{T_o}^{t}}_{t}\, \d r \right] \,\,\,,
\end{equation}
where ${{T_o}^{t}}_{t}$ is the stress-energy tensor of the
solitonic regular solution. This equation can be rewritten as,
\begin{equation}
\label{be2} E_{\rm bind}= - \left[ \int^{\infty}_{r_\Delta} r^2
({T}^{t}_{t}- {{T_o}^{t}}_{t})\, \d r -\int^{r_\Delta}_{0} r^2
{{T_o}^{t}}_{t} \d r \right] \,\,\,.
\end{equation}
Here we can identify the first term as the difference between the
hair of the BH and the ``hair" of the soliton. Of course we are
comparing the quantities (the integrands) that live on different
manifolds, but the total integral is well defined. What happens in
the EYM case is that both stress tensors behave very much alike,
for the exterior region ($r>r_\Delta$) and for large values of
$r_\Delta$, and thus the only term that contributes is the second
one, that gives the ADM mass of the soliton (recall that $r_\Delta
\gg 1$, that is for BH's much larger than the characteristic size
of the soliton (of order one in this dimension-less units), so the
integral captures most of the soliton mass). In the scalar field
case, the fact that the binding energy is proportional to
$r_\Delta$, for large black holes, is captured by the fact that
the BH contribution to the first term in (\ref{be2}) is
dominating. It would be interesting to explore this issue in other
gravity-matter systems.

\section{Instability and final state}
\label{sec:5}


\noindent The next question we want to consider has to do with the
following situation. Consider the case where a hairy black hole of
geometrical radius $r_\Delta$ is slightly perturbed and therefore
it decays. The final state will be, one expects, a black hole that
in its near horizon geometry resembles the Schwarzschild solution,
with the scalar field taking the value where the  potential has a
local minima and vanishes. This means that in this process the
``scalar charge" at the horizon, namely the value $\phi_\Delta$
must change. One can make an argument similar to the one in Ref.
\cite{ACS} to conclude that, in that situation, the horizon must
grow in the process and therefore, the available energy to be
radiated can not all be radiated to infinity; part of it must {\it
fall into the black hole}. Let us now recall the estimate for the
upper bound of the total energy to be radiated.

\begin{figure}
  \includegraphics[angle=0,scale=.70]{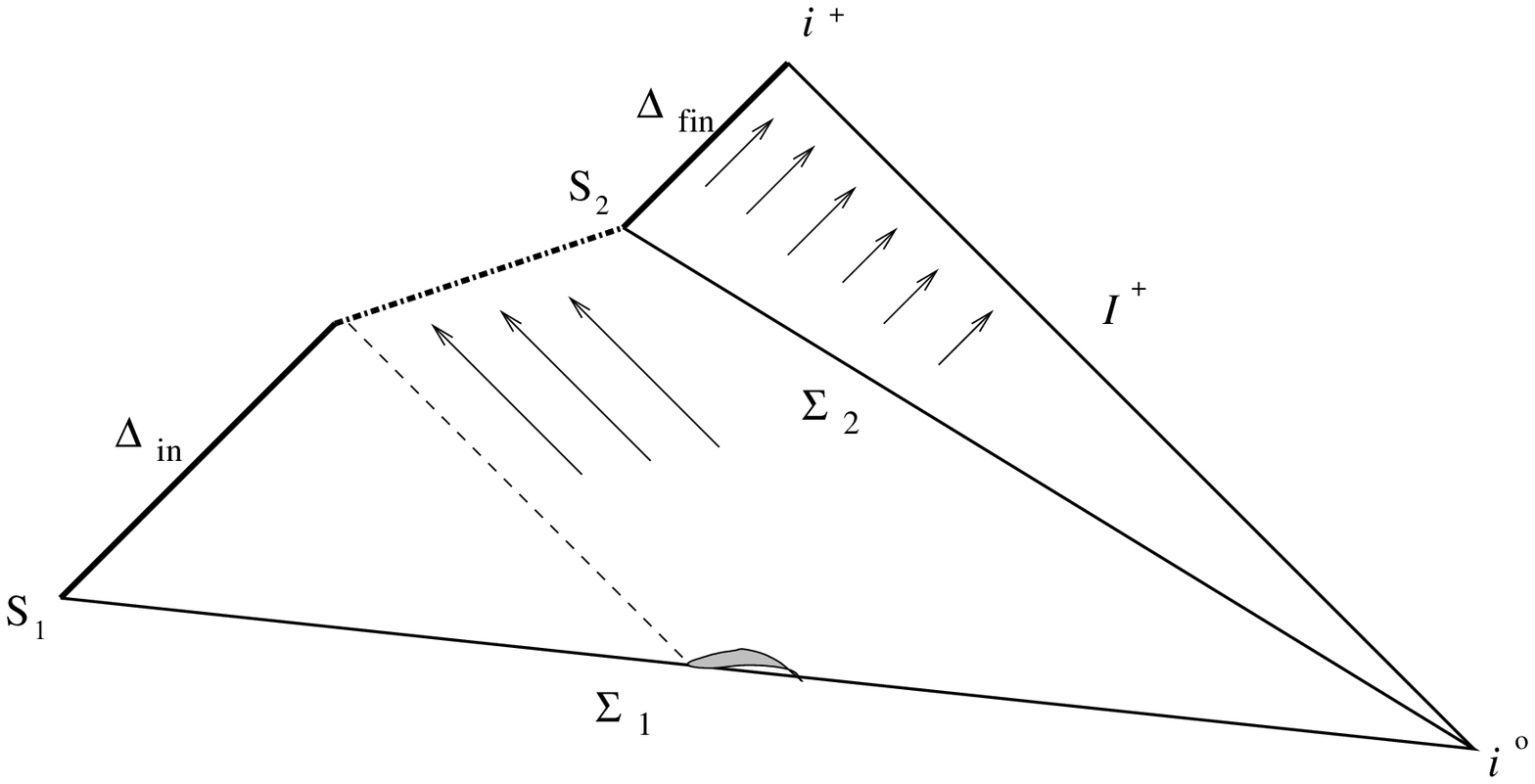}
  \caption{}\label{fig:4} This figure illustrates a physical process where an
  initial configuration with an isolated horizon $\Delta_{\rm in}$ is
  perturbed and the final state contains
  another isolated horizon $\Delta_{\rm fin}$.
\end{figure}

\begin{figure}[h]
\centerline{
 \epsfig{figure=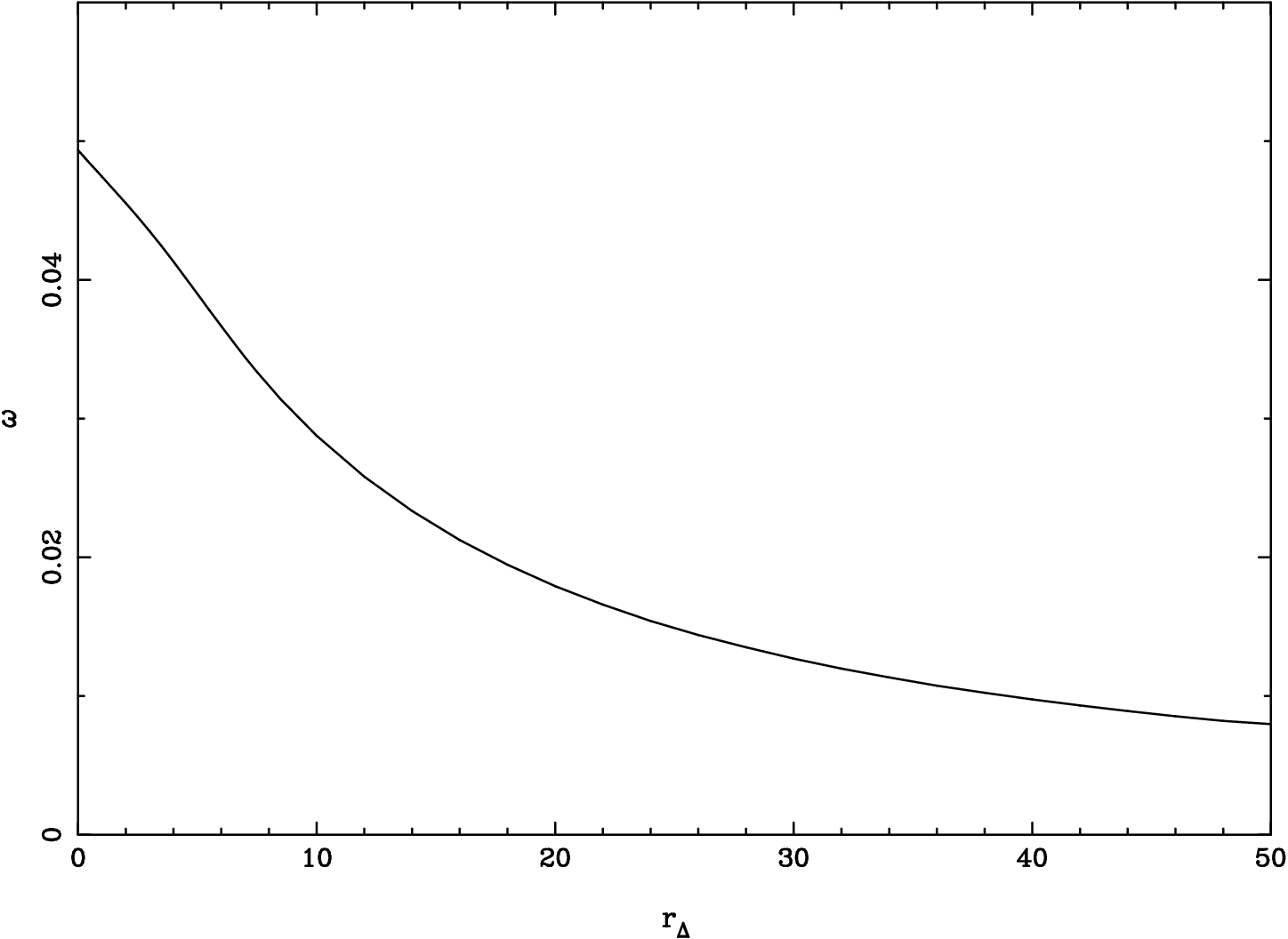,width=7cm,angle=-90}
} \vspace*{0.5cm} \caption{} The unstable-mode frequency
$\omega=\sqrt{-\sigma^2}$ of
the perturbed BH and soliton ($r_\Delta=0$)
is plotted as a function of the horizon
radius $r_\Delta$. Note that the frequency decreases as the
horizon radius increases. \label{Efrec}
\end{figure}

The first step is to assume that the process illustrated in
Fig.~\ref{fig:4} takes place. Then, we assume that in the initial
surface there was an isolated horizon $\Delta_{\rm in}$ and after
the initial unstable configuration has decayed, with part of the
energy falling through the horizon and the rest radiating away to
infinity, we are left with a horizon $\Delta_{\rm fin}$ of a
hairless black hole (with $r_\Delta^{\rm fin}>r_\Delta^{\rm
in}$). If we denote by $E_{{\cal I}^+}$ the energy radiated to
future null infinity ${\cal I}^+$, and given that the ADM energy
does not change in the process, we have
\begin{equation}
 M_{\rm ADM}=M_\Delta(r_\Delta^{\rm in})+M_{\rm sol}=M_\Delta^{\rm
schwarz}(r_\Delta^{\rm fin})+E_{{\cal I}^+}\,,
\end{equation}
which can be rewritten as,
\begin{equation}
 M_{\rm sol}+E^{\rm in}_{\rm bind}=
M_\Delta^{\rm
schwarz}(r_\Delta^{\rm fin})- M_\Delta^{\rm schwarz}(r_\Delta^{\rm
in}) +  E_{{\cal I}^+}\, .
\end{equation}
On the right-hand-side note that the first two terms can be identified
with $\Delta M^{0}_{\Delta}$, namely the change in (bare) horizon
mass, while the second term corresponds to the radiated energy.
Thus, it is natural to identify the quantity on the left as the
{\it available energy} $E_{\rm avail}$ on the system. We can then
write,
\begin{equation}
 E_{\rm avail}= M_{\rm ADM}-
M_\Delta^{\rm schwarz}(r_\Delta^{\rm in})= M_{\rm sol}+E_{\rm bind}
(r^{\rm in}_{\Delta}
)=M_{\rm sol}+E^{\rm new}_{\rm bind}(r^{\rm
in}_{\Delta})+B\,r_{\Delta}^{\rm in}\label{eavai}
\end{equation}
There are several comments regarding this quantity. First, we note
that there is a qualitative change in the behavior of $E_{\rm
avail}(r^{\rm in}_\Delta)$ as function of the initial horizon
radius, as in the EYM case. Its functional dependence is very
similar to the binding energy since they differ only by the
soliton mass. In the EYM case the available energy was equal to
the soliton mass when there was no initial black hole (there is no
energy used in binding the BH), and decreases as the radius
increases. For very large black holes, the available energy goes
to zero. For the scalar case under consideration here, we have a
different behavior. The available energy decreases for small black
holes but starts to increase and grows linearly with $r_\Delta$. The fact that
in the EYM case the available energy went to zero for large BH´s
was interpreted as meaning that those black holes were `less
unstable'. This expectation was confirmed by the fact that the
frequencies of the linear perturbations was decreasing with the
radius of the initial BH \cite{Bizon-Chmaj,ACS}. It is natural
then to ask the same question for the scalar black holes. We have
computed the frequencies of the (single) unstable mode $\psi(t,r)=
\chi(r) e^{\imath \sigma t}$ present (where $\sigma^2$ turns to be
always negative), as a function of the horizon radius and plotted
it in Fig.~\ref{Efrec} (where $\psi(t,r)$ represents a linear
perturbation of $\phi(r)$; see Ref.\cite{SN} for the details).

As can be seen from the figure~\ref{Efrec}, the frequencies still
decrease as the black holes become larger, which is the same
behavior observed in the EYM case. It is convenient then to
reconsider the meaning of `less unstable'. In the scalar field
case considered here, numerical investigations of the dynamical
evolution of the soliton as initial state show that the system is
unstable \cite{evolution}.  The dynamical evolution of the system
depends on the sign of the initial perturbation on the extrinsic
curvature. For one sign of the perturbation, the system collapses
and forms a black hole with a final isolated horizon, while for
the other sign the system expands as a domain wall and gets
therefore radiated to infinity (for details see \cite{evolution}).
One should then expect that the dynamical evolution of slightly
perturbed hairy black holes will show a  similar qualitative
behavior. In that case, for one sign of the perturbation one might
expect the situation considered before, namely that the scalar
field collapses and the BH grows. For the other sign, one can
imagine that there could be, in some situations, an expanding wall
that radiates away while leaving a ``naked black hole". The
pressing question in whether, in that case, this residue would be
a Schwarzschild like or an AdS like black hole. This question
arises  since, for the soliton collapse in the case of the
expanding wall, the region around the origin resembles an AdS
spacetime with an effective negative cosmological constant
generated by the (non-positive) potential. One might need in that
case a new interpretation of the formalism.

It is our belief that one needs to clarify what the criteria
should be for regarding the system as slightly unstable or very
unstable, other than the frequency of its perturbations. This and
a full clarification of the nature of the resulting bare black
hole could be achieved whenever full numerical simulations of
dynamical evolution staring from scalar hairy black holes become
available.

Let us return our discussion to Eq. (\ref{eavai}). The first thing
to note is that due to the characteristic behavior of these
solutions, for horizons larger than $r_\Delta\approx 30$, the
horizon mass of the hairy black hole becomes larger than the
Schwarzschild horizon mass of the same radius. This is also the
point at which the binding energy becomes positive. One can thus
speculate that the black holes of this radius and larger will have
more violent collapses with a larger fraction of the available
energy radiated away. Finally, from Eq. (\ref{eavai}) one could
interpret that again, the term $(B r_\Delta )$ could be associated
to the soliton mass to form a solitonic residue that, together
with the new binding energy allows us to have the same qualitative
features of the heuristic model of \cite{ACS}. Again, a more
detailed analysis will have to wait for the numerical
investigations of the fully dynamical process.

\section{An empirical formula}
\label{sec:6}

\noindent The non-linear behavior of the Einstein-Matter equations
and the non-trivial relations between the masses and the horizon
radius posses a challenge to obtain an analytical formula for
$M_{\rm ADM}(r_\Delta)$. One should expect that there is in general no closed analytical formula for the masses of hairy BH.

We have discovered that the following empirical formula reproduces
the qualitative and quantitative features of the numerical
analysis
\begin{eqnarray}
M_{\rm ADM}^{\rm emp} (r_\Delta) &=& \sqrt{
\left(M_{\rm sol} + \frac{D}{2C} \right)^2 + C^2 r_\Delta^2 +
Dr_\Delta } - \frac{D}{2C} \,\,\,\nonumber
\\
&=& \overline{M_{\rm sol}}\left\{
\sqrt{ \left[1 + \frac{Cr_\Delta}{\overline{M_{\rm sol}}}\right]^2 + F}
 - 1\right\} \,\,\,.
\label{memp}
\end{eqnarray}
where
\begin{equation}
D= \frac{2C\beta_0M_{\rm sol}}{2C-\beta_0}\,\,\,,\,\,\,
\overline{M_{\rm sol}} = \frac{\beta_0 M_{\rm sol}}{2C-\beta_0}
\,\,\,,\,\,\,F= \left(\frac{2C}{\beta_0}\right)^2 -1
\end{equation}
$\beta_0$ being the value of $\beta$ at $r_\Delta=0$.
\bigskip

The formula (\ref{memp}) has the following nice properties:
\bigskip
\begin{enumerate}

\item $M_{\rm ADM}^{\rm emp}(0)= M_{\rm sol}$.

\item $M_{\rm ADM}^{\rm emp} (r_\Delta)$ is a monotonically
increasing
    function of $r_\Delta$.

\item For large $r_\Delta$, $M_{\rm ADM}^{\rm emp}\rightarrow C
r_\Delta$.

\item The relative error between $M_{\rm ADM}^{\rm emp}$ and the
numerical one is less that $10\%$. These errors become very small
for small and large $r_\Delta$.

\item One can define an empirical binding energy by using $E_{\rm
bin}^{\rm emp}(r_\Delta)= M_{\rm ADM}^{\rm emp}- M_{\rm sol} -
Cr_\Delta$ [where the first two terms provide the empirical
horizon mass; here we are using Eq. (\ref{ebindnew}) ]. This formula reproduces very well the numerical
results.

\item One can then obtain a fit for $\beta$ as follows
\begin{equation}
\beta^{\rm em}(r_\Delta) = 2\frac{dM_{\rm ADM}^{\rm emp}}{dr_\Delta}
= \frac{2r_\Delta C^2 + D}{\sqrt{ \left(M_{\rm sol} + \frac{D}{2C}
\right)^2 +
C^2 r_\Delta^2 + Dr_\Delta } }
= \frac{2C\left(1 + \frac{Cr_\Delta}{\overline{M_{\rm sol}}}\right)}
{\sqrt{ \left[1 + \frac{Cr_\Delta}{\overline{M_{\rm sol}}}\right]^2 + F}}
\,\,\,.
\end{equation}
This formula reproduces the qualitative shape of the numerical
$\beta$, such as its exact value at the origin, its monotonically
increasing behavior and its asymptotic value $2C$ for large
$r_\Delta$.

\item For the Schwarzschild case ($C=1/2$, $M_{\rm sol}=0$), one
obtains the expected results: $M_{\rm ADM}^{\rm emp}(r_\Delta)=
r_\Delta/2$,
 $\beta^{\rm em}(r_\Delta) \equiv 1$,
$E_{\rm bind}^{\rm emp}(r_\Delta) \equiv 0$.
\bigskip

\end{enumerate}

\begin{figure}[h]
\centerline{
\epsfig{figure=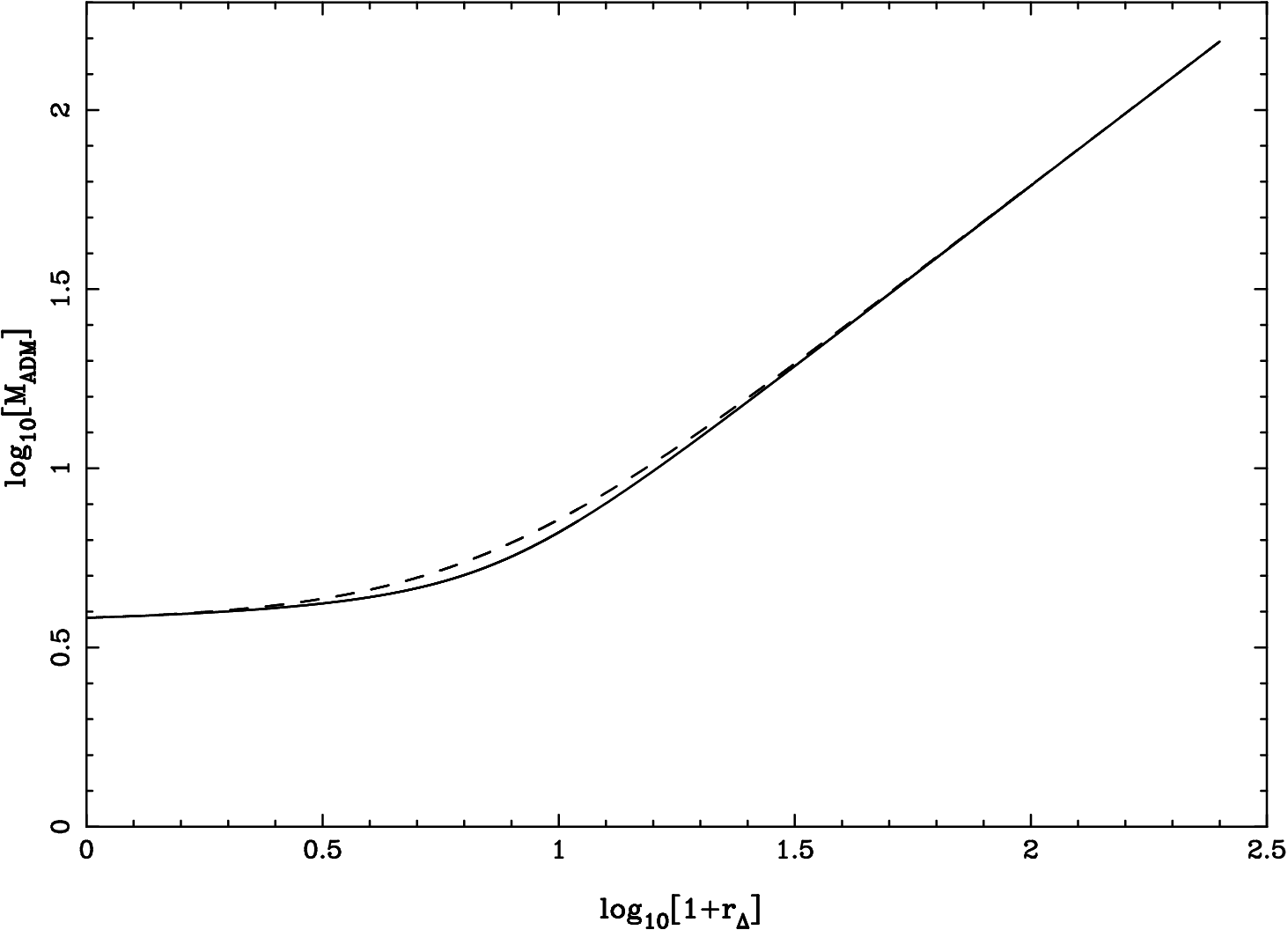,width=4.5cm,angle=-90}
\epsfig{figure=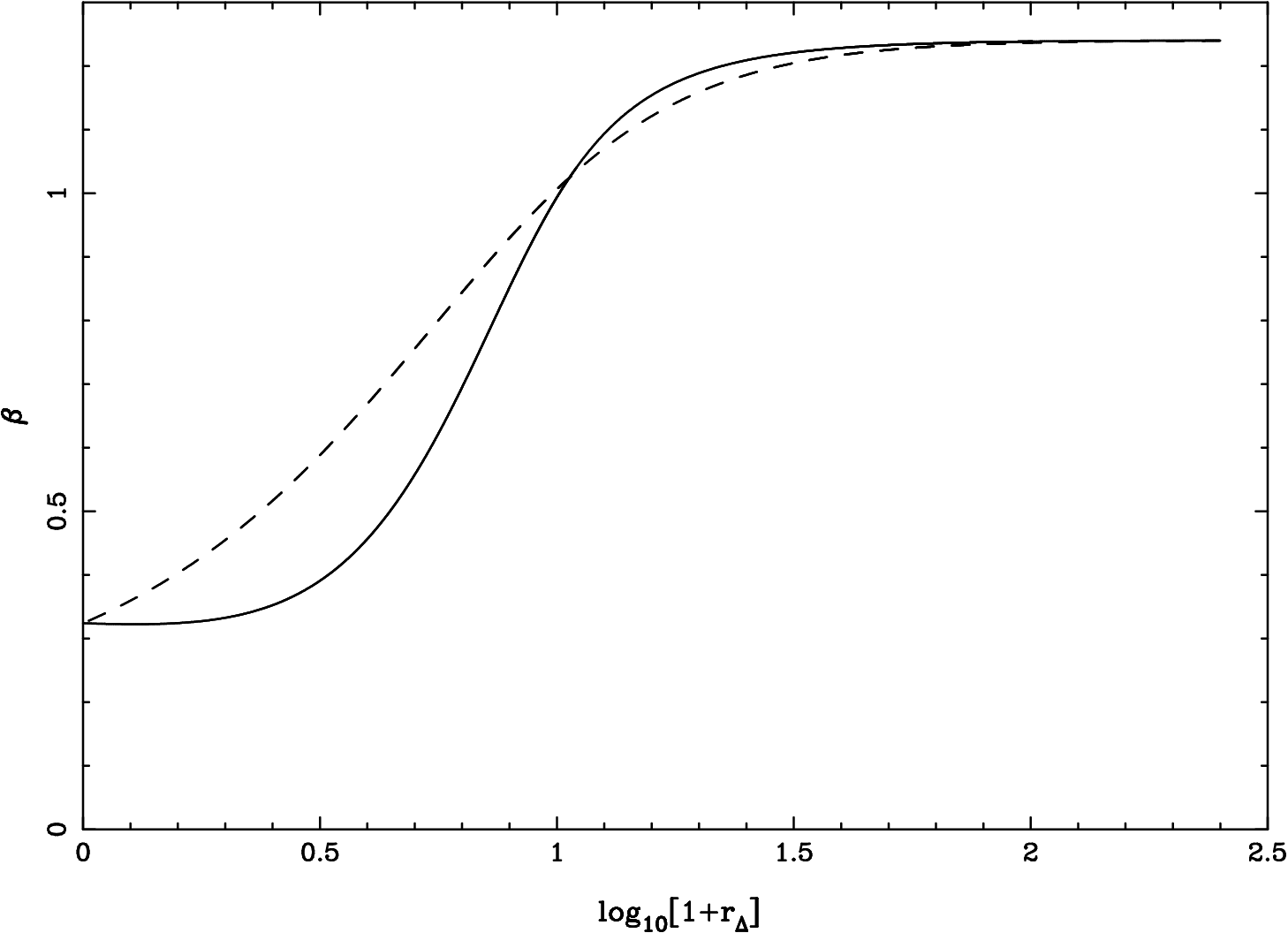,width=4.5cm,angle=-90}
\epsfig{figure=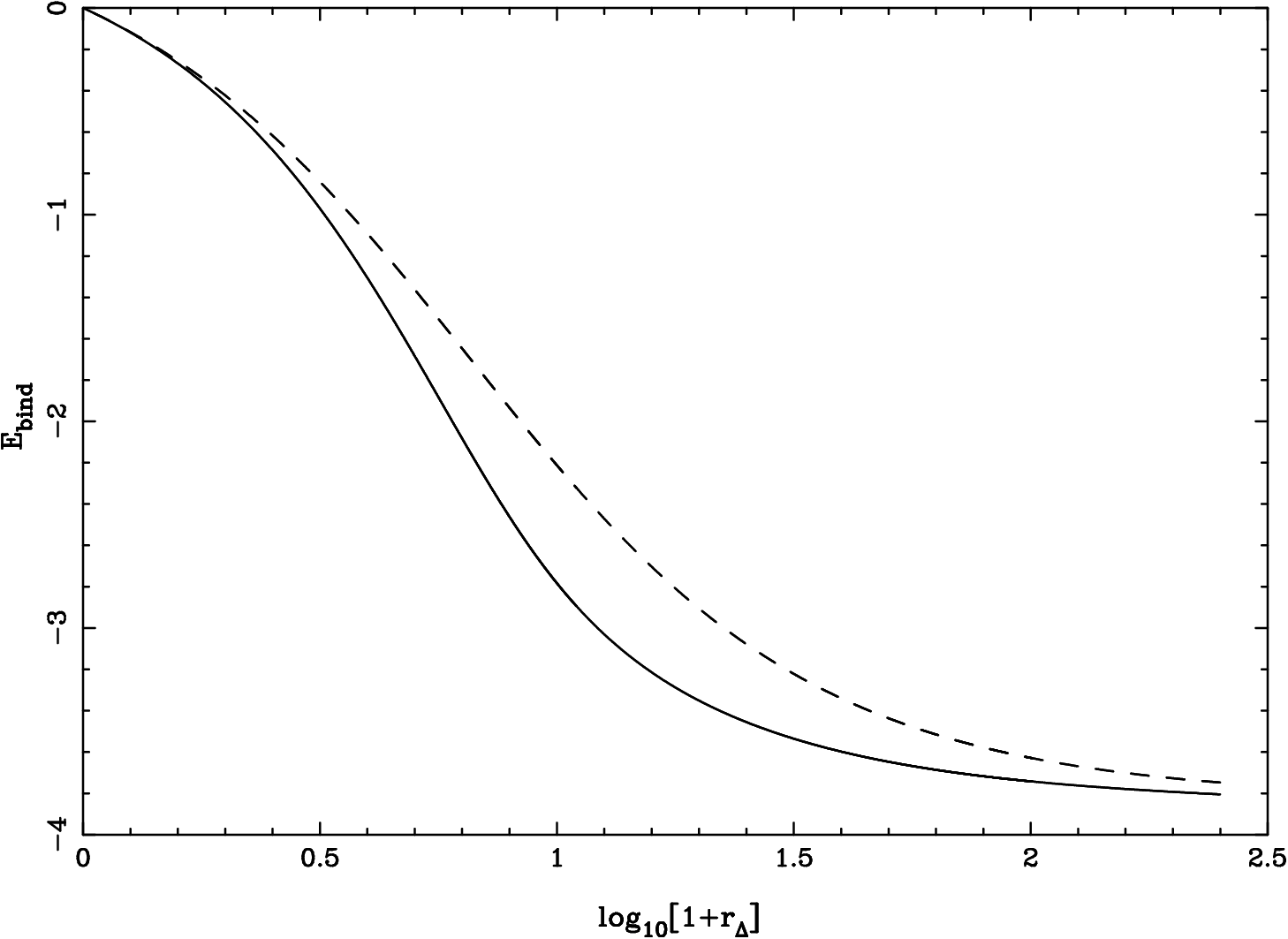,width=4.5cm,angle=-90} }
\vspace*{0.5cm} \caption{}
The three panels depict the ADM-mass,
$\beta$, and binding energy $E_{\rm bind}^{\rm new}$ , respectively, as a function of the
horizon radius. The solid lines correspond to the values obtained
from a numerical analysis and the dashed lines were obtained from
the empirical formulae described in the main text. Note the good
qualitative behavior of the empirical formulae. Remarkably good
fits to the more precise numerical values are obtained for small
and large $r_\Delta$. The values for the empirical formula are
$C\approx 0.62$, $M_{\rm sol}\approx 3.827$ and $\beta_0\approx 0.324$.
\label{fits}
\end{figure}

Clearly by adding terms of the form $r_\Delta^\theta$
($1<\theta<2$) inside the square root of Eq.(\ref{memp}) one could
improve the fit between the numerical results and the analytical
formula. In Figure~\ref{fits} we compare
between the empirical formula and the numerical values of the
hairy scalar black holes, for the ADM mass, screened surface
gravity $\beta$ and the binding energy.

\begin{figure}[h]
\centerline{
\epsfig{figure=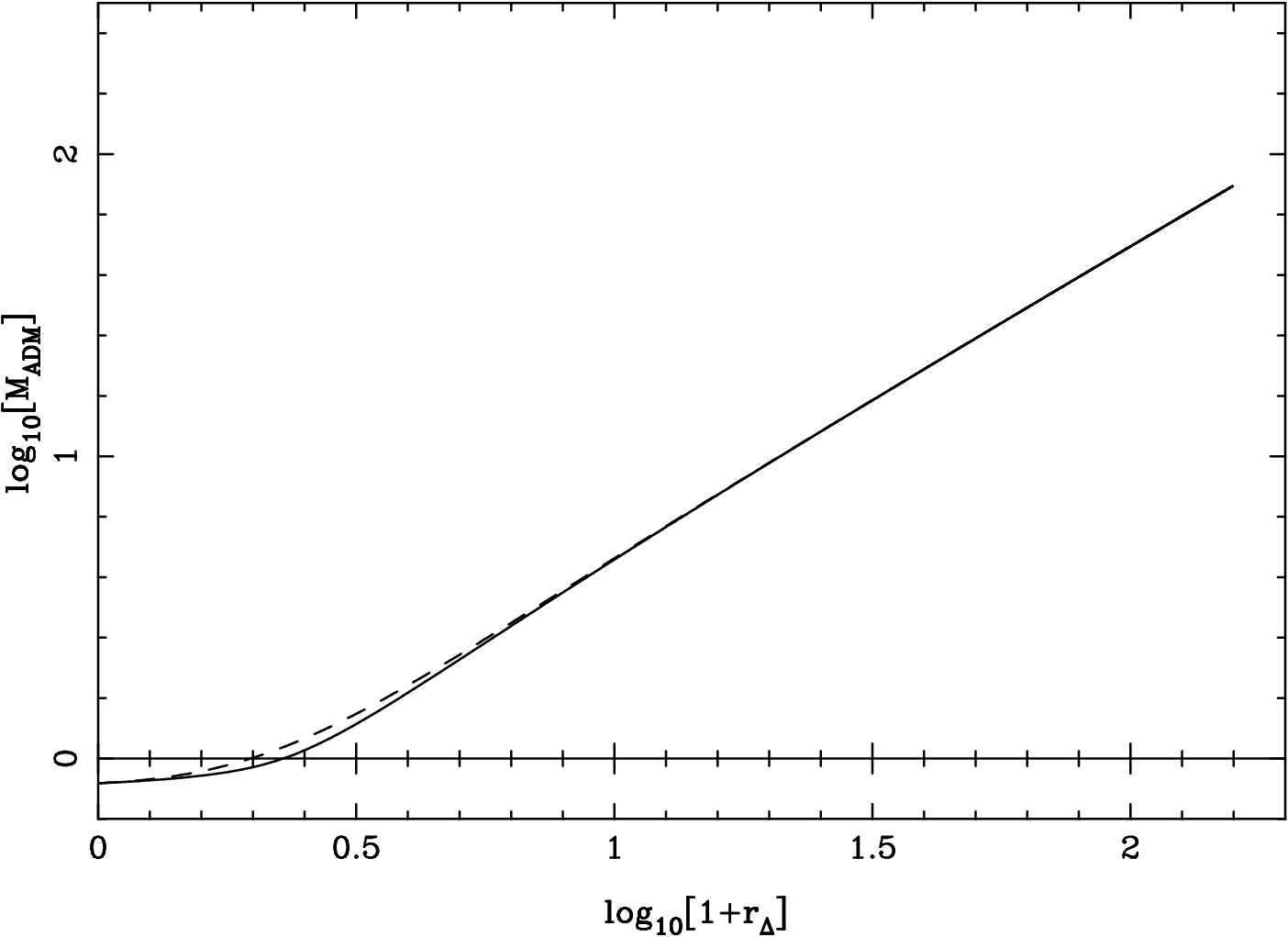,width=4.5cm,angle=-90}
\epsfig{figure=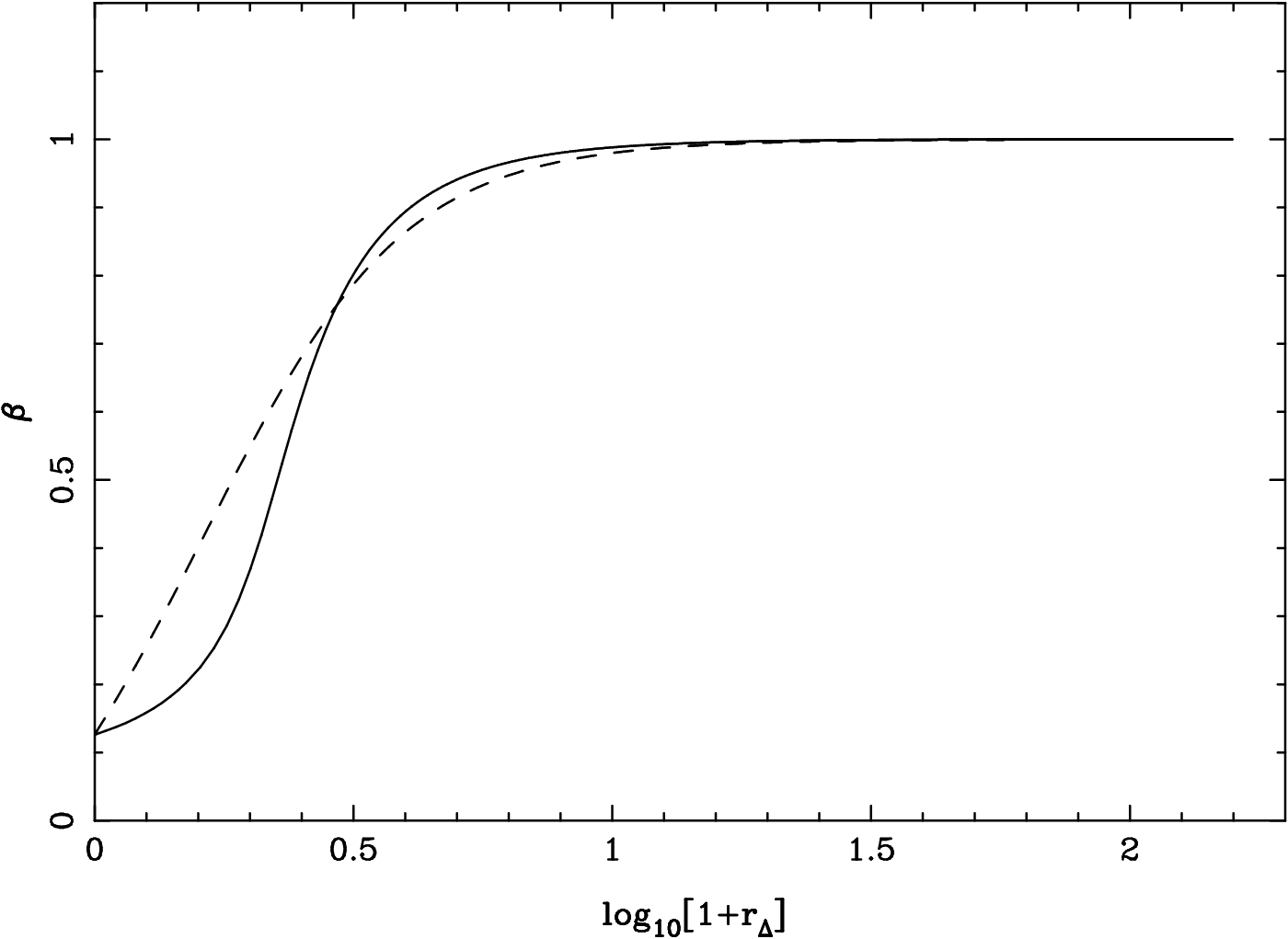,width=4.5cm,angle=-90}
\epsfig{figure=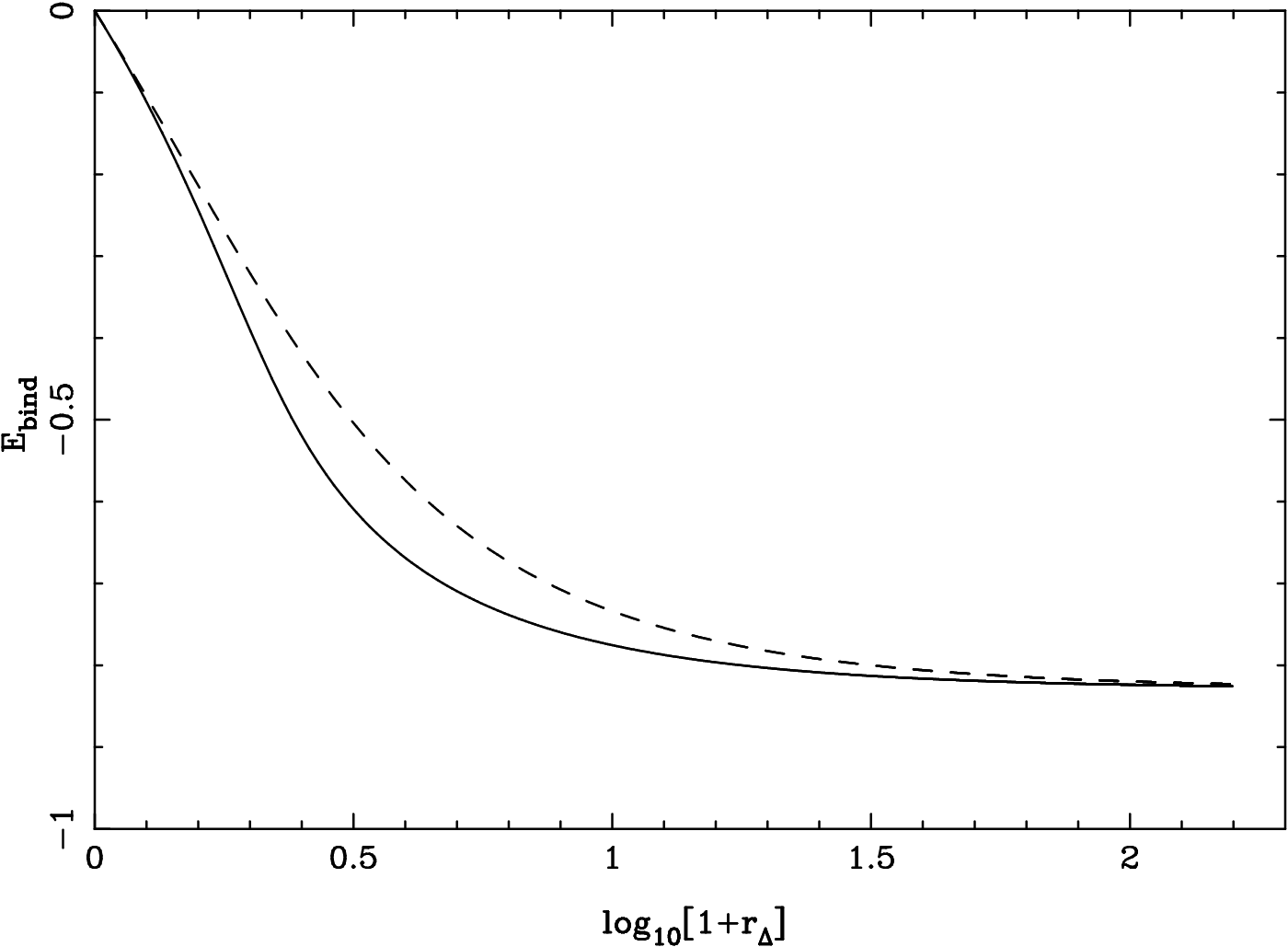,width=4.5cm,angle=-90} }
\vspace*{0.5cm} \caption{} Same as Fig.\ref{fits}, for the
Einstein-Yang-Mills theory ($n=1$ colored black holes). Here the
values for the empirical formulae are $C=1/2$, $M_{\rm
sol}\approx 0.828$ and $\beta_0\approx 0.126$. \label{fitseym}
\end{figure}

The empirical formula can be used also for the EYM case with
$C=1/2$ and the corresponding values of $M_{\rm sol}$ and $D$.
Figure~\ref{fitseym} compares the numerical values of the EYM
$n=1$-branch with those obtained from the empirical formula.

We conjecture that the empirical formula can work also for different $n$,
by using their corresponding values $M_{\rm sol}^{n}$, and $\beta_0^{n}$.
Moreover, we also speculate that such a formula can hold for other theories
admitting hair, such as in the Einstein-Skyrme and Einstein-sphaleron models.
It remains to be investigated what are the values of $C$, $M_{\rm sol}$ and
$\beta_0$ for such cases.
\bigskip

Now, we can further use the first law of thermodynamics
$\delta M= \kappa \delta A_\Delta/(8\pi)$ for
$M_{\rm ADM}^{\rm emp}$, and obtain the following prediction
\begin{equation}
\label{pred}
\kappa\left(M_{\rm ADM}^{\rm emp} + \frac{D}{2C}\right)= C^2 + \frac{D}{2r_\Delta}
\,\,\,.
\end{equation}
Since the properties 1)$-$7) show that the analytical results
obtained from Eq. (\ref{memp}) work particularly well for large
and small $r_\Delta$, the most reliable consequence of
(\ref{pred}) is a remarkable simple relation between the surface
gravity and the ADM mass for sufficiently large hairy black holes:
\begin{equation}
\label{pred2}
\kappa M_{\rm ADM}\approx C^2
\,\,\,.
\end{equation}

Note that Eq. (\ref{pred2}) is consistent for the
Schwarzschild case ($C=1/2$, $D=0$, $M_{\rm ADM}=r_\Delta/2$), where
the identity $\kappa M_{\rm ADM}\equiv 1/4$, holds exactly.

\bigskip

We have performed a non-exhaustive analysis of the solutions
with respect to variations of some of the parameters of the scalar
potential Eq. (\ref{potential}). Notably, we have computed the effect
of the variation of $\eta_2$ on the global quantities. It is to note that
changing $\eta_2$ modifies
the potential barrier between the global and the local minimum.
In fact, the closer the value $\eta_2$ to $\eta_1/2$, the less
negative is $V(a+\eta_1)= \lambda \eta_1^3\left[2\eta_2 -\eta_1\right]/12$,
and therefore the potential approaches the conditions where the
no-hair theorems apply. The details of the solutions then depend in a
non-trivial fashion between the interplay of the
negative global
minimum (in order to avoid the applicability of the non-hair theorems)
and the height of the potential barrier.

\begin{figure}[h]
\centerline{
\epsfig{figure=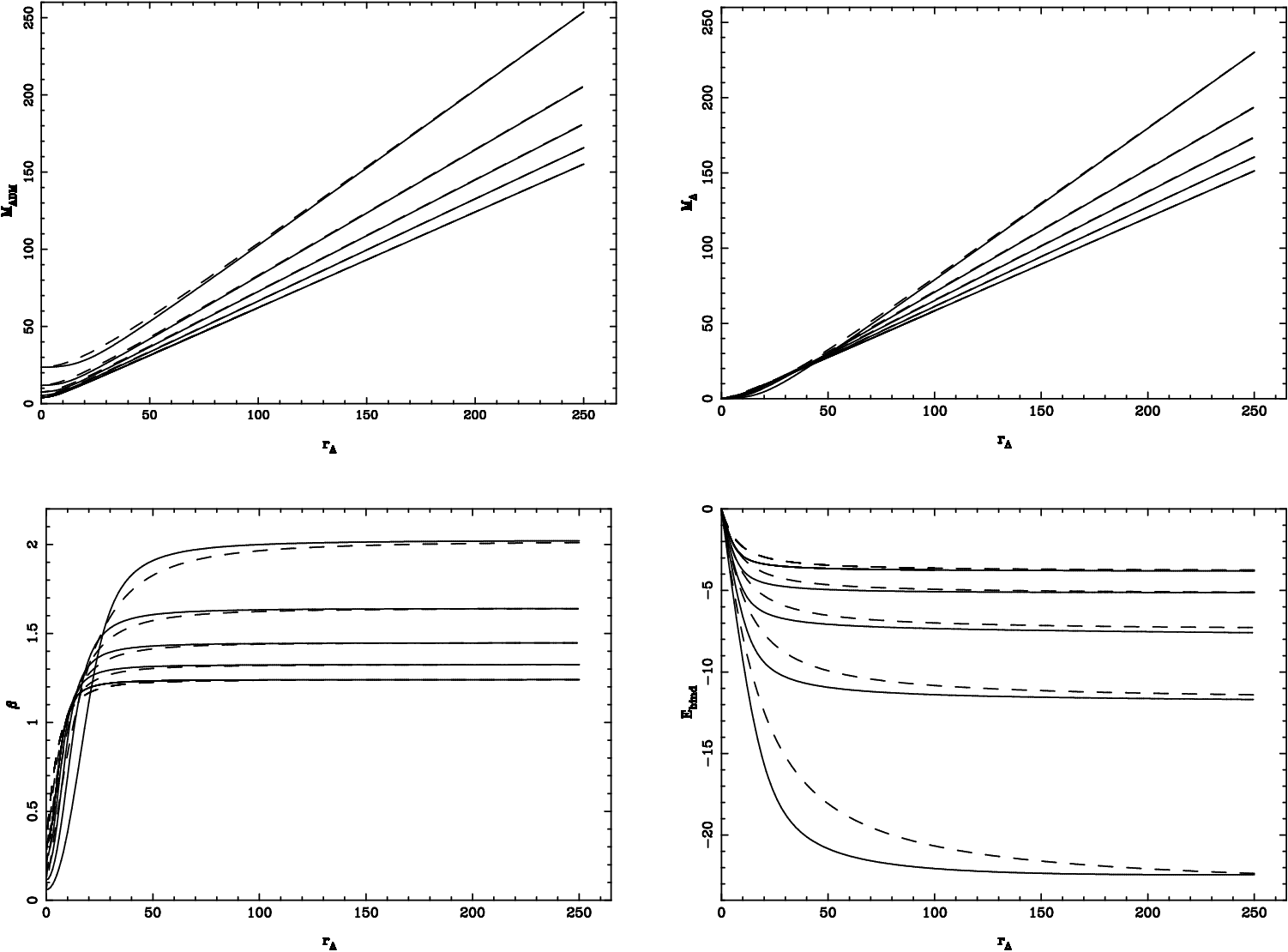,width=10.4cm,angle=-90} }
\vspace*{0.5cm} \caption{}
Panels 1-4 depict the ADM-mass, horizon mass, $\beta$ and the binding energy
respectively as a function of $r_\Delta$. The solid lines were obtained
from the numerical analysis while the dashed lines were computed using the
empirical formulae. The lines are associated with the five different values used for
$\eta_2= 0.1,0.11,0.12,0.13,0.14$ with $\eta_1=0.5$ fixed.
As seen from bottom to top (for large $r_\Delta$) the plots of panels 1-3 correspond to
$\eta_2$ in increasing order (in panel 4 the order is reversed).
The values of the parameters $\vec{P}_{\eta_2}=
\left(M_{\rm sol}, C,\beta_0\right)_{\eta_2}$
used in the empirical formulae are
$\vec{P}_{0.1}\approx
\left(3.82,0.62,0.32\right)$,
$\vec{P}_{0.11}\approx
\left(5.22,0.66,0.24\right)$,
$\vec{P}_{0.12}\approx
\left(7.47,0.72,0.17\right)$
$\vec{P}_{0.13}\approx
\left(11.78,0.82,0.11\right)$
$\vec{P}_{0.14}\approx
\left(23.51,1.01,0.06\right)$.
 \label{multgraf}
\end{figure}

Figure ~\ref{multgraf} depicts different global quantities as a
function of the horizon radius for five different values of
$\eta_2$. The soliton mass ($r_\Delta=0$) as well as the ADM and
horizon masses (for large $r_\Delta$) tend to increase with
$\eta_2$. Remarkably, the empirical formulae continue to provide
reasonable good results by changing the corresponding values of
their parameters $\vec{P}_{\eta_2}:= \left(M_{\rm sol},
C,\beta_0\right)_{\eta_2}$. The quality of the fit to the numerical values
can be appreciated by the dashed curves of Fig. ~\ref{multgraf}
which were computed with the empirical formulae.

\section{Discussion}
\label{sec:7}
\bigskip

\noindent Let us first summarize our results. By solving numerically
Einstein's equations for static solutions of a self-gravitating
scalar field, we have analyzed the behavior of several spacetime
quantities as functions of the black hole horizon radius. We have
found that the ADM mass of the spacetimes exhibits two types of
behavior: it is  similar to other `hairy' theories for small black
holes, but its behavior changes dramatically for large black holes.
In particular the ADM mass of large BH scales not as $r_\Delta/2$ as
in other theories (EYM, EYMH, etc), but the proportionality constant
(with respect to horizon radius) takes a different value depending
on the form of the potential ($C\approx 0.63$ for $\eta_2=0.1$). In
this article we have analyzed the consequences of this fact for a
model based on the isolated horizons formalism. In such a model, a
hairy black hole is viewed as a bound state of a soliton (which we
have) and a `bare black hole'. The binding energy is found to be
negative in EYM and EYMH, but in our case, for large BH, the binding
energy becomes positive and grows linearly with $r_\Delta$ \footnote{Recently,
another system in 5 dimensions was shown to posses a positive
binding energy as well \cite{hart}.}. This fact leads to several
possibilities. We have seen that it is possible to modify the
original model by `renormalizing' the binding energy in such a way
that the newly defined energy has the same qualitative behavior as
in the EYM system. The price one has to pay is the need to
reinterpret either a new `solitonic residue', or a new bare black
hole. As a first attempt towards giving a definite answer to this
question, we analyzed the frequency of the unstable mode of the
linearized perturbation, and found that the behavior is the same as
in EYM. This suggests that the proper physical interpretation is
still unclear and that further numerical dynamical investigations
are needed to fully settle the question. In particular, the two
different regimes of the theory might have some consequences in the
dynamical evolution of slightly perturbed BHs, where one could
conjecture a different qualitative behavior for small and large
black holes, regarding the endpoint of evolution and the nature of
the {\it bare} black hole to which the solution settles. We have
also conjectured that the constant that fixed the proportionality
between ADM mass and horizon radius for large BH's is a
theory-dependent constant, which would in particular imply that
axi-symmetric non-spherical BH solutions to the gravity-scalar field
system would have the same asymptotic behavior, for each given
potential. It would be worth studying other gravity-matter systems,
such as non-minimally coupled scalars, to see whether they posses a
different proportionality constant (work is in progress in these
directions).

We have shown also that a very simple heuristic analytic formula
captures the essential qualitative behavior of the ADM mass of the
hairy scalar BH's, specially for small and large values of the
horizon radius. We have conjectured that such formula can also be
useful for EYM and more general hairy black holes. It remains a
theoretical challenge to fully understand the origin of such simple formula.

Perhaps the most important conclusion from the present work is the
lesson that hairy black holes for different matter systems exhibit
new, and sometimes, unexpected behavior. This also point out to
the need of a proper and deeper understanding of the reason  why
the heuristic hairy black hole model works so well for the system
that it does, and whether the phenomenological modifications we
have proposed here stand the test of full numerical
investigations.





\section*{Acknowledgments}

We would like to thank A. Ashtekar and D. Sudarsky for discussions.
This work was in part supported by grants DGAPA-UNAM IN122002,
and IN119005. U.N. acknowledges partial support from SNI, and Grants No. 4.8
CIC-UMSNH, No. PROMEP PTC-61 and No. CONACYT 42949-F. 














\begin{thebibliography}{99}
\bibitem{comment}
Another example of hairy BH's with the property that the masses
$M_{\rm hair}$ scale also linearly with $r_\Delta$ are the hairy
BH's of the Einstein-Skyrme system. To see this we take the
equations of motion for the Einstein-Skyrme model as described in
the subsection (7.4) from the reference \cite{volkov}; specifically,
we take the equation (7.37):
\begin{eqnarray}
\frac{dm(r)}{dr}&=& N \left[ \frac{r^2}{2} + \sin^2 \chi \right]
\left(\frac{d\chi}{dr}\right)^2 + \left[ r^2 + \frac{\sin^2 \chi}{2}
\right] \frac{\sin^2 \chi}{r^2} \label{7.37}
\end{eqnarray}
where the metric for static and spherically symmetric configurations
is:
\begin{eqnarray}
ds^2 &=& -\sigma^2(r) N(r) dt^2 + \frac{1}{N(r)} dr^2 + r^2
d\Omega^2
\end{eqnarray}
with
\begin{eqnarray}
N(r) = 1 - 2m(r)/r,
\end{eqnarray}
and the Skyrmion field $\chi(r)$ depending on the dimensionless
coordinate $r$; to continuation we define a new coordinate as
$x=r/r_\Delta$ and after integration rewrite (\ref{7.37}) as
\\
\begin{eqnarray}
M_{\rm ADM}(r_\Delta) = \frac{r_\Delta}{2} &+& M_{\rm
hair}(r_\Delta) \nonumber \\
=\frac{r_\Delta}{2} &+& r_\Delta \int^\infty_1 \left[\frac{N x^2}{2}
\left(\frac{d\chi}{dx}\right)^2 + \sin^2 \chi \right] dx
\nonumber \\
&+& \frac {1}{r_\Delta} \int^\infty_1 \left[N
\left(\frac{d\chi}{dx}\right)^2 + \frac{\sin^2 \chi}{2x^2} \right]
(\sin^2 \chi) dx \label{4}
\end{eqnarray}
The behavior for $\chi(x)$ can see from left panel in the fig. 14.
of such reference. From equation (7.41) we see the asymptotic
behavior for $\chi(x)$ as $x \rightarrow \infty$ (with $a$ a
constant): $\chi \approx ax^{-2}$. Altough this analysis shows that 
$M_{\rm ADM}(r_\Delta)$ grows linearly with $r_\Delta$ for $r_\Delta\gg 1$, 
we can not apply the
limit $r_\Delta \rightarrow \infty$ to (\ref{4}) because the
existence of Skyrme's black holes is limited to configurations with
horizon radius $r_\Delta \lesssim r_\Delta^{max}(\kappa)$ (here
$\kappa\equiv 4\pi G_0 f^2$ is the coupling constant of the theory
and $r_\Delta^{max}(\kappa)$ is a maximal value depending of
$\kappa$) and $\kappa \lesssim \kappa_{bh}^{max}$. The value of
$\kappa_{bh}^{max}$ is of the order of $0.0315$. At the other hand,
the existence of solitons is permitted for $\kappa \lesssim 0.0437$. 
In the Eisntein-Higgs at hand, $r_\Delta$ seems to be limited by the precision 
of the shooting method. For large $r_\Delta$, the shooting parameter 
$\phi_\Delta$ approaches a below limit ($\approx 0.26$ for 
the model $\vec{P}_{0.1}$ of fig. \ref{multgraf}). It is unknown if this limit 
for the SHBH configurations is a fundamental limit or not. 
It turns then interesting to construct a similar model based in the 
isolated-horizon formalism for the Einstein-Skyrme system to compare 
similarities and differences with the model presented here \cite{Nielsen}.
\end{thebibliography}
\end{document}